\documentclass[twocolumn, english, reprint, superscriptaddress, breaklinks=true, showkeys, showpacs=false]{revtex4}

\usepackage[T1]{fontenc}
\usepackage{color}
\usepackage{babel}
\usepackage{float}
\usepackage{amssymb}
\usepackage{listings}
\usepackage{graphicx}
\usepackage{amsmath}
\usepackage{svg}
\usepackage{tabularx}
\usepackage{algorithm2e}
\usepackage{algpseudocode}

\usepackage[unicode=true,pdfusetitle, bookmarks=true,bookmarksnumbered=false,bookmarksopen=false, breaklinks=true,pdfborder={0 0 0},backref=false,colorlinks=true]{hyperref}
\usepackage{breakurl}
\definecolor{Code}{rgb}{0,0,0}
\definecolor{Decorators}{rgb}{0.5,0.5,0.5}
\definecolor{Numbers}{rgb}{0.5,0,0}
\definecolor{MatchingBrackets}{rgb}{0.25,0.5,0.5}
\definecolor{Keywords}{rgb}{0,0,1}
\definecolor{self}{rgb}{0,0,0}
\definecolor{Strings}{rgb}{0,0.63,0}
\definecolor{Comments}{rgb}{0,0.63,1}
\definecolor{Backquotes}{rgb}{0,0,0}
\definecolor{Classname}{rgb}{0,0,0}
\definecolor{FunctionName}{rgb}{0,0,0}
\definecolor{Operators}{rgb}{0,0,0}
\definecolor{Background}{rgb}{0.98,0.98,0.98}
\lstdefinelanguage{Python}{
numbers=left,
numberstyle=\footnotesize,
numbersep=1em,
xleftmargin=1em,
framextopmargin=2em,
framexbottommargin=2em,
showspaces=false,
showtabs=false,
showstringspaces=false,
frame=l,
tabsize=4,
basicstyle=\ttfamily\small\setstretch{1},
backgroundcolor=\color{Background},
commentstyle=\color{Comments}\slshape,
stringstyle=\color{Strings},
morecomment=[s][\color{Strings}]{"""}{"""},
morecomment=[s][\color{Strings}]{'''}{'''},
morekeywords={import,from,class,def,for,while,if,is,in,elif,else,not,and,or,print,break,continue,return,True,False,None,access,as,,del,except,exec,finally,global,import,lambda,pass,print,raise,try,assert},
keywordstyle={\color{Keywords}\bfseries},
morekeywords={[2]@invariant,pylab,numpy,np,scipy},
keywordstyle={[2]\color{Decorators}\slshape},
emph={self},
emphstyle={\color{self}\slshape},
}

\makeatletter
\@ifundefined{textcolor}{}
{%
 \definecolor{BLACK}{gray}{0}
 \definecolor{WHITE}{gray}{1}
 \definecolor{RED}{rgb}{1,0,0}
 \definecolor{GREEN}{rgb}{0,1,0}
 \definecolor{BLUE}{rgb}{0,0,1}
 \definecolor{CYAN}{cmyk}{1,0,0,0}
 \definecolor{MAGENTA}{cmyk}{0,1,0,0}
 \definecolor{YELLOW}{cmyk}{0,0,1,0}
}

\hypersetup{colorlinks=true,citecolor=blue,linkcolor=cyan,urlcolor=blue,filecolor= green, breaklinks=true}
\usepackage{url}

\makeatother

\begin{document}

\title{Learning to learn with an evolutionary strategy \\ applied to variational quantum algorithms }

\author{Lucas Friedrich\href{https://orcid.org/0000-0002-3488-8808}{\includegraphics[scale=0.05]{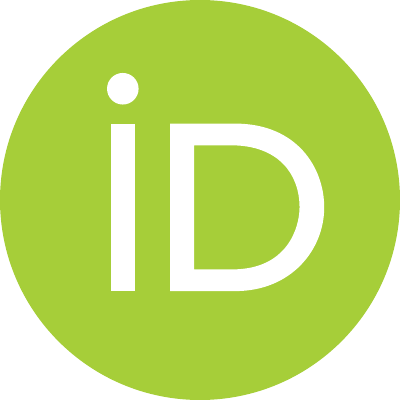}}}
\email{lucas.friedrich@acad.ufsm.br}
\affiliation{Physics Department, Center for Natural and Exact Sciences, Federal University of Santa Maria, Roraima Avenue 1000, Santa Maria, RS, 97105-900, Brazil}

\author{Jonas Maziero\href{https://orcid.org/0000-0002-2872-986X}{\includegraphics[scale=0.05]{orcidid.pdf}}}
\email{jonas.maziero@ufsm.br}
\affiliation{Physics Department, Center for Natural and Exact Sciences, Federal University of Santa Maria, Roraima Avenue 1000, Santa Maria, RS, 97105-900, Brazil}


\begin{abstract}
Variational Quantum Algorithms (VQAs) employ parameterized quantum circuits optimized using classical methods to minimize a cost function. While VQAs have found broad applications, certain challenges persist. Notably, a significant computational burden arises during parameter optimization. The prevailing ``parameter shift rule'' mandates a double evaluation of the cost function for each parameter. In this article, we introduce a novel optimization approach named “Learning to Learn with an Evolutionary Strategy” (LLES). LLES unifies “Learning to Learn” and “Evolutionary Strategy” methods. “Learning to Learn” treats optimization as a learning problem, utilizing recurrent neural networks to iteratively propose VQA parameters. Conversely, “Evolutionary Strategy” employs gradient searches to estimate function gradients.
Our optimization method is applied to two distinct tasks: determining the ground state of an Ising Hamiltonian and training a quantum neural network. The obtained results underscore the efficacy of this novel approach. Additionally, we identify a key hyperparameter that significantly influences gradient estimation using the ``Evolutionary Strategy'' method.
\end{abstract}

\keywords{Variational quantum algorithms, Quantum neural networks, Evolutionaty Strategy, Learning to Learn}

\maketitle

\section{Introduction}

Variational quantum algorithms (VQAs) represent a leading approach for harnessing the potential of Noisy Intermediate-Scale Quantum (NISQ) devices. In VQAs, a parameterization denoted as $U$ is constructed by the combination of various quantum gates, some of which are parametrized by $\pmb{\theta}$. Classical optimization techniques are then employed to iteratively update these parameters, aiming to minimize a cost function denoted as $C$. VQAs have already demonstrated their versatility in addressing a wide array of problems. These include solving systems of linear equations \cite{linear_system}, simulating quantum systems \cite{quantum_simulation}, estimating the Fisher information \cite{BECKEY}, and applications in quantum neural networks \cite{MANGINI, FARHI}, among others.

Despite the diverse applications of VQAs, certain challenges remain unresolved, with one significant obstacle known as barren plateaus (BPs) gaining special attention \cite{McClean,MLaroccaBP}. In practice, the optimization of VQA parameters relies on gradients of the cost function. However, as the size of the parameterized quantum circuit $U$ grows, the gradient of the cost function often diminishes, rendering the optimization of parameters $\pmb{\theta}$ highly challenging. Several factors have been linked to the emergence of BPs, including the choice of the cost function \cite{BR_cost_Dependent}, entanglement \cite{BR_Entanglement_devised_barren_plateau_mitigation, BR_Entanglement_induced_barren_plateaus}, noise \cite{BR_noise}, and circuit expressibility \cite{BR_expressibility}. Remarkably, a study by Ref. \cite{BR_gradientFree} highlighted that this issue is also prevalent in gradient-free optimization methods. 
Numerous approaches have been proposed to mitigate BPs, such as strategies involving initialization \cite{Friedrich_BP,BR_initialization_strategy}, inducing large gradients through correlations \cite{BR_Large_gradients_via_correlation}, Long Short-Term Memory networks (LSTM) \cite{BR_LSTM}, and layer-wise optimization \cite{BR_layer_by_layer}. Nonetheless, this remains an active and evolving research area.

Another crucial challenge associated with Variational Quantum Algorithms (VQAs) revolves around the construction of the parameterization $U$. As this parameterization is derived from the composition of various quantum gates, the number of potential combinations grows exponentially with the size of $U$. This raises a fundamental question: What constitutes the optimal approach for constructing $U$? Answering this question necessitates consideration of multiple factors.
For instance, one key factor is expressibility. Research has established a correlation between expressibility and the performance of quantum neural network models \cite{Hubregtsen}. It is observed that greater expressibility often leads to improved model performance. However, an intriguing counterpoint is presented in Ref. \cite{BR_expressibility}, indicating that highly expressive parameterizations are more susceptible to the challenges posed by barren plateaus (BPs), impeding effective training and, consequently, reducing performance. Moreover, as explored in Ref. \cite{Friedrich_parametrization_expressivity}, highly expressive parameterizations can result in cost functions that concentrate around a fixed value.
Besides, the architecture of the quantum chip was shown to not significantly impact the performance of the parameterization $U$ \cite{Friedrich_architecture}. Given the complexities arising from these and other factors, various studies have proposed methods to automate the process of constructing $U$ \cite{Zhang, Du_Yuxuan, Kuo}. These efforts aim to provide systematic solutions to the intricate challenge of parameterization in VQAs.

Another significant challenge that VQAs confront is the formidable computational expense involved in optimizing their parameters. Optimization is facilitated by a classical optimizer that typically employs the gradient of the cost function for parameter updates. The commonly used method for gradient computation is the `parameter shift rule' \cite{Crooks_parameters_shift_rule, Schuld_parameters_shift_rule}. This method is based on the expression:
\begin{equation}
    \frac{\partial C}{\partial \theta_{k}} = \frac{1}{2}\big( C(\theta_{k}+\pi/2) - C(\theta_{k}-\pi/2) \big).
    \label{eq:derivada_C}
\end{equation}
This equation is utilized to calculate partial derivatives of the cost function, and subsequently, the gradient. In the general context, the cost function is defined as:
\begin{equation}
    C(\pmb{\theta}) := \langle H \rangle = \mathrm{Tr}[HU\rho U^{\dagger}].\label{eq:cost_1}
\end{equation}
Here, $H$ represents an observable, and $\rho$ is the initial state, often defined as $\rho := V|0\rangle \langle0|V^{\dagger}$, with $V$ serving as a classical data encoding parameterization. Depending on the application, Eq. \eqref{eq:cost_1} can further serve as input to another function. In quantum machine learning, for example, the goal is to build a model that, when provided with an input $x$, produces the output $y$. In such cases, the result derived from Eq. \eqref{eq:cost_1} is commonly employed as the model's output, and the aim is to optimize the parameters so that this output aligns closely with the desired result.
Moreover, parameterization $V$ is employed to encode data $x$ into a quantum state. As expressed in Eq. \eqref{eq:derivada_C}, computing the derivatives entails executing the quantum circuit twice for each parameter, leading to a linear growth in the number of circuit executions required to obtain the gradient. Several methods have been proposed to address this challenge \cite{Friedrich_es, Anand_es, Rebentrost, Sweke, bowles2023backpropagation}, yet it remains an active area of study.

In this article, our primary objective is to introduce a novel optimization method that draws inspiration from two established approaches: `Learning to Learn' \cite{Andrychowicz} and `evolutionary strategies' \cite{Friedrich_es, Anand_es}. Our motivation is rooted in the observation that techniques relying solely on the gradient of a function often neglect valuable information that could enhance model performance. Notably, in the context of VQAs, essential information such as the degree of entanglement and the nature of observables tends to be overlooked by gradient-based methods.
While some prior works, such as Ref. \cite{BR_LSTM}, have proposed the utilization of the `Learning to Learn' method to optimize VQA parameters, this method also relies on Eq. \eqref{eq:derivada_C} for gradient computation, resulting in substantial computational costs. To confront this challenge, we propose a synergistic approach. By integrating the `Learning to Learn' method with `evolutionary strategies,' we aim to estimate the gradient of the cost function more efficiently. Our goal is to markedly reduce the computational demands required for VQA parameter optimization.

The subsequent sections of this article are structured as follows: 
In Sec. \ref{sec:method}, we introduce the optimization method we propose. We commence by providing concise overviews of the `Learning to Learn' method (Sec. \ref{subsec:LL}) and the `Evolutionary Strategy' method (Sec. \ref{subsec:es}). Subsequently, in Sec. \ref{subsec:LLES}, we present a detailed description of the novel method we are introducing here.
In Sec. \ref{sec:Applications}, we explore two practical applications of our newly proposed method. In Sec. \ref{subsec:Ground_state}, we scrutinize its performance in ground state energy estimation. Following that, in Sec. \ref{subsec:qnn}, we investigate its behavior in the training of a quantum neural network.
To conclude our study, we provide a summary of our key findings in Sec. \ref{sec:Conclusion}.

\section{Method}
\label{sec:method}

In this section, we introduce our new optimization method, entitled `Learning to Learn with an Evolutionary Strategy' (LLES). LLES integrates the Learning to Learn (LL) method with the Evolutionary Strategy (ES), both widely recognized in their respective domains. First, we will provide a brief description of these individual methods - LL and ES - and then proceed to a detailed explanation of the LLES method.

\subsection{Learning to Learn}\label{subsec:LL}

The `Learning to Learn' method employs a recurrent neural network, particularly a Long Short Term Memory (LSTM) network, to generate parameters for a quantum circuit. 
An LSTM model is an advanced recurrent neural network architecture designed to handle data with temporal correlations. Developed to overcome the limitations of traditional RNNs, such as the vanishing gradient problem, the LSTM uses a sequence of cells that control the flow of information through three main gates: the input gate $i_{t}$, the forget gate $f_{t}$, and the output gate $o_{t}$.

The input gate determines which new information will be stored in the cell state $C_{t}$. To achieve this, it first decides which values should be updated. This decision is described by:
\begin{equation}
    i_t = \sigma(W_{ii}x_t + b_{ii} + W_{hi}h_{t-1} + b_{hi}),
\end{equation}
where $\sigma(.)$ is the sigmoid function. Next, a vector of new candidate values to be added to the cell state is created through the equation:
\begin{equation}
    \widetilde{C}_{t} = \tanh(W_{ig}x_t + b_{ig} + W_{hg}h_{t-1} + b_{hg}),\label{eq:estadoDaCelula}
\end{equation}
where $\tanh(.)$ is the hyperbolic tangent.

The forget gate, in turn, decides which old information from the cell state should be discarded. This process is described by the equation:
\begin{equation}
    f_{t} = \sigma(W_{if}x_{t} + b_{if} + W_{hf}h_{t-1} + b_{hf}).
\end{equation}
After that, the cell state is updated using the expression:
\begin{equation}
    C_t = f_t \odot C_{t-1} + i_t \odot \widetilde{C}_{t},\label{eq:estadoDaCelula}
\end{equation}
where $\odot$ represents element-wise multiplication (or Hadamard product). 

Finally, the output gate regulates the hidden state $h_{t}$ of the LSTM, determining which information from the cell will be used in the next step. This process is described by the following equations:
\begin{equation}
    o_t = \sigma(W_{io}x_t + b_{io} + W_{ho}h_{t-1} + b_{ho}),
\end{equation}
and  
\begin{equation}
    h_t = o_t \odot \tanh(C_t).
\end{equation}

This structure allows the model to efficiently store and access relevant data, even in long sequences, significantly improving performance on tasks involving complex temporal dependencies.

Originally, the LL method was developed to address classical optimization problems \cite{Andrychowicz}. However, in Ref. \cite{BR_LSTM}, its application was extended to encompass challenges in the domain of quantum problems.

\begin{figure}
    \centering
    \includegraphics[scale=0.47]{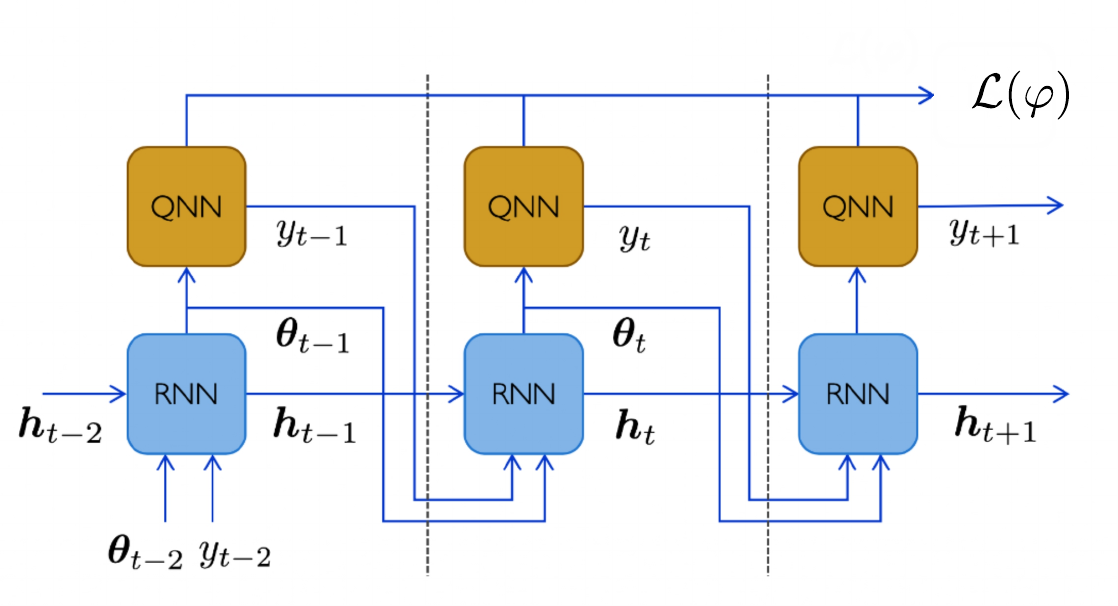}
    \caption{This figure illustrates the operational concept of the `Learning to Learn' method. The cost function is a function of the values of $y_{t'}$. This figure is adapted from Ref. \cite{BR_LSTM}.}
    \label{fig:LL_model}
\end{figure}

In conventional practice, parameter optimization adheres to the following update rule:
\begin{equation}
    \pmb{\theta}^{t+1} = \pmb{\theta}^{t} - \eta \nabla C.
    \label{eq:otimization_rule}
\end{equation}
Here, $t$ represents the epoch number, $\eta$ is the learning rate, and $C$ denotes the cost function. However, in `Learning to Learn' (LL), the optimization rule itself becomes a task to be learned. Consequently, we reformulate this rule as:
\begin{equation}
    \pmb{\theta}^{t+1} = \pmb{\theta}^{t} +  g(C, \varphi).\label{eq:novaRegra}
\end{equation}
Here, $g(C, \varphi)$ is a function dependent on the cost function and the parameters $\varphi$ that need to be learned. As we can observe in Eq. \eqref{eq:estadoDaCelula}, the expression used to update the cell state, after the appropriate manipulations, becomes similar to Eq. \eqref{eq:novaRegra}, which describes our new optimization rule. Thus, this is achieved through the use of an LSTM network, which iteratively proposes parameter updates for the quantum circuit, referred to as the Quantum Neural Network (QNN), according to Ref. \cite{BR_LSTM}. The LSTM network initially receives as input the parameters $\pmb{h}_{t-2}$, $\pmb{\theta}_{t-2}$, and $y_{t-2}$, usually computed as $y \sim \langle H \rangle$, and returns the parameters $\pmb{h}_{t-1}$ and $\pmb{\theta}_{t-1}$. The parameters $\pmb{\theta}_{t-1}$ are used as QNN parameters to generate $y_{t-1}$. Subsequently, the new parameters $\pmb{h}_{t-1}$, $\pmb{\theta}_{t-1}$, and $y_{t-1}$ are fed as input to the LSTM network, which again generates new values for $\pmb{h}$ and $\pmb{\theta}$. This process is repeated for a total of $T$ iterations, and finally, using the values of $y_{i}$ with $i = t-2, t-1, t, t+1,...$, the LSTM computes the cost function $\mathcal{L}$. For a visual representation of this method, refer to Fig. \ref{fig:LL_model}.

The training process involves the optimization of parameters $\varphi$ within the LSTM network. This optimization adheres to the same rule as presented in Eq. \eqref{eq:otimization_rule}, but in this case, it pertains to the parameters $\varphi$. As depicted in Fig. \ref{fig:LL_model}, this optimization necessitates the computation of gradients for the QNN layer. Moreover, given that the method entails $T$ interactions between the LSTM network and the QNN layer, gradient computation for the QNN layer must be performed a total of $T$ times.
If we consider that the quantum layer depends on $p$ parameters and utilize Eq. \eqref{eq:derivada_C} for gradient calculation of the QNN layer, the quantum circuit must be executed a total of $2pT$ times to perform the optimization of parameters $\varphi$. Consequently, it is evident that this optimization approach incurs a substantial computational cost. For more comprehensive details on this method, refer to Refs. \cite{Andrychowicz, BR_LSTM}.

In the following subsection, we will introduce the `Evolutionary Strategy' method, which leverages a gradient search technique to more efficiently estimate the gradient of a function $f$.

\subsection{Evolutionary Strategy}\label{subsec:es}

In the `Evolutionary Strategy' method (ES), we employ a gradient search technique to estimate the gradient of a given function $f$. Specifically, for a function $f(\pmb{z})$ with $\pmb{z} \in \mathbb{R}^{m}$ and a probability distribution $\pi(\pmb{z},\pmb{\theta})$, we define:
\begin{equation}
    J(\pmb{\theta}) = \int f(\pmb{z})\pi(\pmb{z},\pmb{\theta})d\pmb{z}. \label{eq:es}
\end{equation}
Calculating the gradient of Eq. \eqref{eq:es} with respect to $\pmb{\theta}$, we obtain:
\begin{equation}
    \nabla_{\pmb{\theta}}J(\pmb{\theta}) = \mathbb{E}_{\pmb{\theta}}[ f(\pmb{z})  \nabla_{\pmb{\theta}} \log \pi(\pmb{z},\pmb{\theta}) ]. \label{eq:es_1}
\end{equation}
From Eq. \eqref{eq:es_1}, we can estimate the gradient using the samples $\pmb{z}_{1}, \ldots, \pmb{z}_{\lambda}$ as:
\begin{equation}
    \nabla_{\pmb{\theta}}J(\pmb{\theta}) \approx \frac{1}{\lambda} \sum_{k=1}^{\lambda}f(\pmb{z}_{k})\nabla_{\pmb{\theta}}\log\pi(\pmb{z}_{k},\pmb{\theta}). \label{eq:es_2}
\end{equation}
Here, the gradient is estimated using $\lambda$ samples, typically defined as:
\begin{equation}
\lambda : = 4 + 3\log(p),\label{eq:lambda}
\end{equation}
where $p$ signifies the number of parameters of the VQA. It is worth mentioning that $\lambda$ must be an integer, as it is used in the summation of Eq. \eqref{eq:es_2}, so we must always use the closest integer given by Eq. \eqref{eq:lambda}. This gradient estimation method, as described in Eq. \eqref{eq:es_2}, is commonly known as the `Canonical Search Gradient.' Moreover, alternative methods for gradient estimation using a gradient search include the `Exponential Natural Evolution Strategies' and the `Separable Natural Evolution Strategies.' For further details on the derivation of Eq. \eqref{eq:es_2} and information about these related techniques, refer to Ref. \cite{Wierstra}.

Since the gradient estimate in Eq. \eqref{eq:es_2} depends on the value of $\lambda = 4 + 3\log(p)$, it is evident that the number of times the quantum circuit must be executed to estimate the gradient scales logarithmically with the number of parameters, in contrast to the linear scaling with Eq. \eqref{eq:derivada_C}. Consequently, the computational cost of obtaining the gradient using Eq. \eqref{eq:es_2} is significantly lower than that using Eq. \eqref{eq:derivada_C}, particularly when dealing with a large number of parameters.

Following Refs. \cite{Salimans,Trust_region}, if we now consider a normal distribution $\pi(\mathbf{z}_k,\pmb{\theta})=\mathcal{N}(\pmb{\theta},\sigma^2\mathbb{I})$, the gradient can be estimated using the equation:
\begin{equation}
    \nabla_{\pmb{\theta}}J(\pmb{\theta}) = \frac{1}{2\lambda \sigma^{2}} \sum_{k=1}^{\lambda} [ f(\pmb{z}_{k}) - f(2\pmb{\theta}-\pmb{z}_{k}) ](\pmb{z}_{k} - \pmb{\theta}).\label{eq:es_3}
\end{equation}
A relevant observation regarding this gradient estimate concerns the choice of the normal distribution. In general, in studies employing the ES method, the analytical expression for the gradient estimate, as presented in Eq.\eqref{eq:es_3}, is derived under the assumption of a normal distribution. However, one might question whether this choice is indeed the most suitable. Could there be another distribution that yields better results? It is likely that, for each specific problem, there exists a more appropriate distribution: in some cases, it will be the normal distribution itself; in others, it may be a different alternative.

The equation (\ref{eq:es_3}) will be used for gradient estimation in our work. Below is shown the pseudocode for obtaining the gradient estimate using this equation.

\begin{algorithm}[H]
    \caption{Example of pseudocode used to estimate the gradient using the Eq. \eqref{eq:es_3}. }
    \SetAlgoLined
    \KwIn{$f(.)$, $\pmb{\theta}$,$\sigma$}
    \KwOut{$\nabla J(\pmb{\theta})$}
    
    \SetKwFunction{Funcao}{Função}
    
    \SetKwProg{Fn}{Função}{}{}
    
    \For{$k = 1, 2, \dots, \lambda $}{
            $ \pmb{z}_{k} \sim  \mathcal{N}(\pmb{\theta}, \sigma^{2}\mathbb{I}) $\;
            $f(\pmb{z}_{k})$\;
            $f(2\pmb{\theta}-\pmb{z}_{k})$\;
        }
        $ \nabla J(\pmb{\theta}) = \frac{1}{2\lambda \sigma^{2}} \sum_{k=1}^{\lambda} [ f(\pmb{z}_{k}) - f(2\pmb{\theta}-\pmb{z}_{k}) ](\pmb{z}_{k} - \pmb{\theta})$\;
    \Return{$\nabla J(\pmb{\theta})$}\;
\end{algorithm}

\subsection{Learning to Learn with Evolutionary Strategy}
\label{subsec:LLES}

The method we introduce, called `Learning to Learn with Evolutionary Strategy' (LLES), combines the principles of LL and ES methods. In other words, we use the LL method to transform the optimization rule for the parameters of a quantum circuit into a learning problem and employ the ES method to efficiently estimate the gradient of the quantum circuit required for optimizing the parameters of the LSTM layer. By using the ES method to estimate the quantum circuit gradient, we reduce the computational cost, which previously scaled linearly with the number of circuit parameters, to a cost that scales logarithmically, as shown in Table \ref{tb:metodo_custo}.

Additionally, as demonstrated in Ref. \cite{BR_Large_gradients_via_correlation}, by correlating the parameters used by the quantum circuit, we can avoid the BPs problem. For example, consider the following cost function:
\begin{equation}
    C(\pmb{\theta}) = \langle \psi |H| \psi \rangle,\label{eq:loss_function_gs}
\end{equation}
where \( H = \mathbb{I} - |0\rangle \langle 0|^{\otimes n} \) and \( |\psi \rangle = \mathcal{M}(\pmb{\theta})|0\rangle^{\otimes n} \), with
\begin{equation}
    \mathcal{M}(\pmb{\theta}) = \bigg( \prod_{i=1}^L e^{-i \frac{\theta_{i}}{2}\sigma_{y}} \bigg)^{\otimes n}.\label{eq:M_bp}
\end{equation}
The variance of the partial derivative of the cost function \( C \) with respect to a parameter \( k \) will be approximately \( \mathrm{Var}[\partial_{k}C] \sim \mathcal{O}(\sqrt{n}) \), meaning it will not suffer from the BPs problem.

In the proposed method, the parameters used by the quantum circuit are correlated through the parameters of the LSTM layer, i.e., \( \theta = \theta(\varphi) \). According to Ref. \cite{BR_Large_gradients_via_correlation}, this implies that the new method also does not suffer from the BPs problem. However, as we can observe in Eq. \eqref{eq:M_bp}, the correlation of parameters exploited in Ref. \cite{BR_Large_gradients_via_correlation} is a direct correlation, meaning that a given parameter is used by different rotation gates. For instance, the parameter \( \theta_{0} \) is applied in the first rotation gate acting on all qubits of the circuit.

On the other hand, in the proposed method, the parameter correlation is indirect, which implies that, in general, the parameters used by different rotation gates are not identical. Therefore, although there is an indirect correlation between the parameters generated by the LSTM layer, this correlation does not guarantee that the BPs problem is avoided with this new method. In fact, as shown in Ref. \cite{Friedrich_parametrization_expressivity}, even if the cost function gradient is not used to optimize the quantum circuit parameters, the high expressiveness of the circuit induces cost function concentration, thus hindering training.

\begin{table}[H]
\setlength{\tabcolsep}{4pt} 
\renewcommand{\arraystretch}{1.2} 
\begin{tabularx}{\linewidth}{|l|>{\centering\arraybackslash}X|}
\hline
\textbf{Method} & \textbf{Computational cost} \\ \hline
\textbf{GRAD} & 2p \\ \hline
\textbf{LL} & 2pT \\ \hline
\textbf{LLES} & 2(4+3log(p))T \\ \hline
\end{tabularx}
\caption{The table presents the computational cost of each method, measured by the number of quantum circuit executions required to optimize the parameters. GRAD represents the standard optimization method, which uses the gradient calculated by Eq. \eqref{eq:derivada_C}. LL corresponds to the \textit{Learning to Learn} method, which also uses Eq. \eqref{eq:derivada_C} to compute the gradients. LLES, on the other hand, refers to the LL method but with gradient estimation performed via ES. In this context, $p$ represents the number of parameters to be optimized, while $T$ indicates the number of interactions between the recurrent layer and the quantum layer in the LL and LLES methods.}
\label{tb:metodo_custo}
\end{table}

\section{Applications}
\label{sec:Applications}

In this section, we demonstrate the effectiveness of the LLES method through two distinct applications: ground state estimation for a given Hamiltonian \cite{vqe} and binary classification using quantum neural networks. It is important to note that the primary aim of this article is to introduce our novel optimization method rather than delve into extensive discussions of ground state estimation and quantum neural networks.

\subsection{Ground state energy estimation}\label{subsec:Ground_state}

\begin{figure}[t]
    \centering
    \includegraphics[scale=0.31]{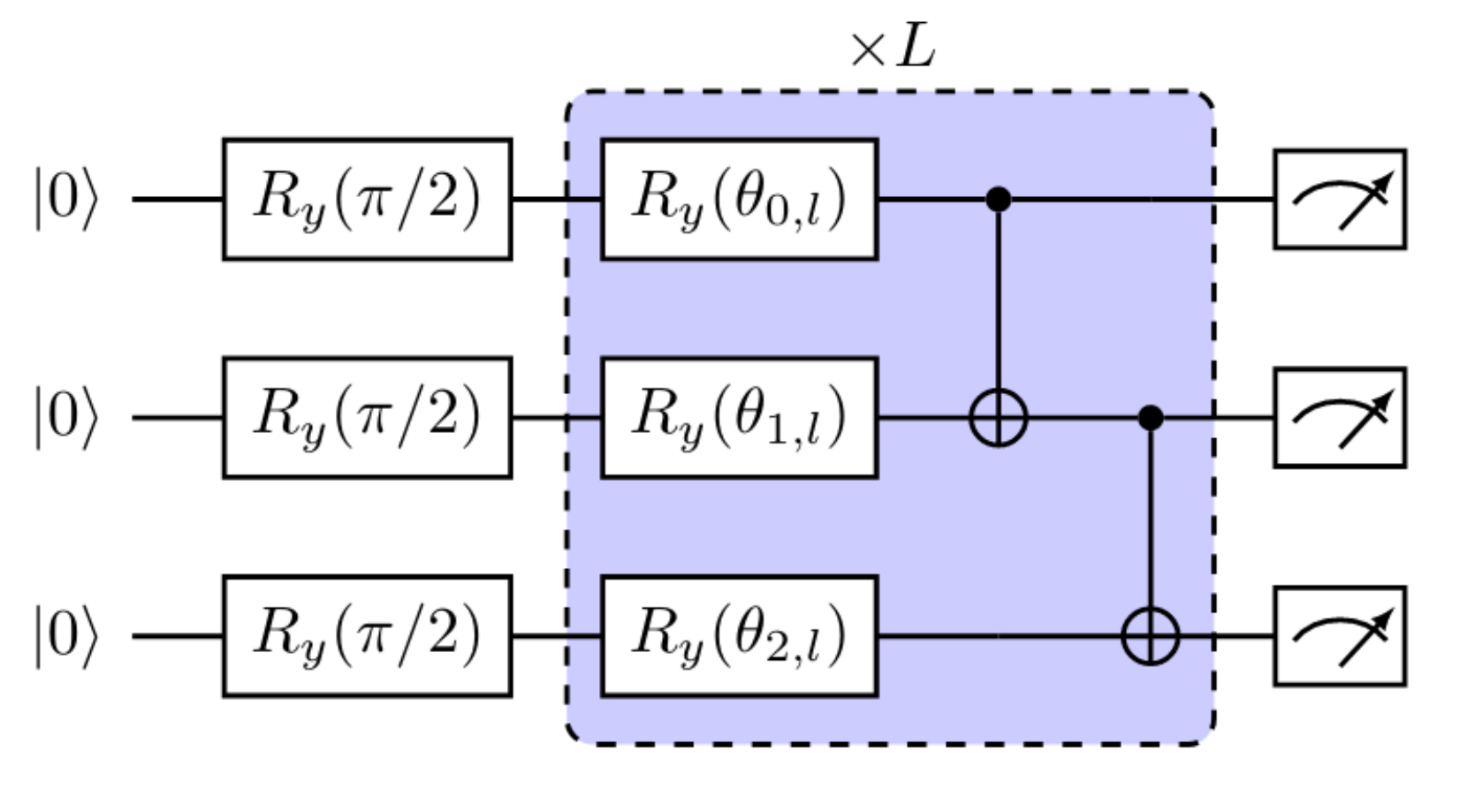}
    \caption{
    Illustration of the quantum circuit used in our simulations. The circuit begins with the preparation of a state by applying the \(R_{y}\) rotation gate with an angle of \(\pi/2\) to all qubits. Subsequently, a parameterization, obtained as the product of \(L\) layers, is applied. Each of the \(L\) layers is constructed identically, as depicted in the illustration, and relies on the optimization of parameters \(\pmb{\theta}\).}
    \label{fig:model_circuit_gs}
\end{figure}

In this subsection, we apply the LLES method to the ground state energy estimation problem. Specifically, we focus on obtaining the ground state for a Hamiltonian $H,$ which is defined as:
\begin{equation}
 H = \bigotimes_{j=1}^{n} \sigma_{Z}^{j},
 \end{equation}
where $n$ represents the number of qubits, and each \(\sigma_{Z}^{j}\) corresponds to one of the Pauli matrices acting on the state space of the $j$-th qubit. 
In this problem, our goal is to find the parameters \(\pmb{\theta}\) that minimize the cost function defined in Eq. \eqref{eq:loss_function_gs} and where $|\psi\rangle$ is obtained using the quantum circuit illustrated in Figure \ref{fig:model_circuit_gs}. However, in this new method, the cost function that we must minimize is defined as:
\begin{equation}
    L(\varphi) = \frac{1}{T}\sum_{j=1}^{T}w_{j}C_{j}. \label{eq:newLossFunction}
\end{equation}
Here \(C_{j}\) represents the \(j\)-th output of the Quantum Neural Network (QNN) layer, calculated using Eq. \eqref{eq:loss_function_gs}, and \(w_{j}\) is a weighting factor, indicating the influence of the \(j\)-th output on the cost function. For the following results, we set \(w_{j} = 1\) for all \(j\). 

To assess the effectiveness of the LLES method, we compare its performance with standard optimization, which we refer to as `GRAD'. This method involves updating the parameters $\pmb{\theta}$ using the rule shown in Eq.  \eqref{eq:otimization_rule}, with the gradient obtained through Eq. \eqref{eq:derivada_C}, where the cost function is defined in Eq. \eqref{eq:loss_function_gs}. Additionally, we compare the LLES method with the standard LL method, where the gradient is also obtained from Eq.  \eqref{eq:derivada_C}, and the cost function is defined in Eq. \eqref{eq:newLossFunction}.

In all three scenarios, the choice of the learning rate (\(\eta=lr\)) is crucial. To evaluate the impact of various \(lr\) values, we consider three distinct settings: \(lr=0.1\), \(lr=0.01\), and \(lr=0.001\).
Additionally, the LLES method relies on a hyperparameter, \(\sigma\), and we investigate three different values: \(\sigma = \frac{\pi}{6}\), \(\sigma = \frac{\pi}{12}\), and \(\sigma = \frac{\pi}{24}\). 

\begin{figure*}
    \centering
    \includegraphics[scale=0.39]{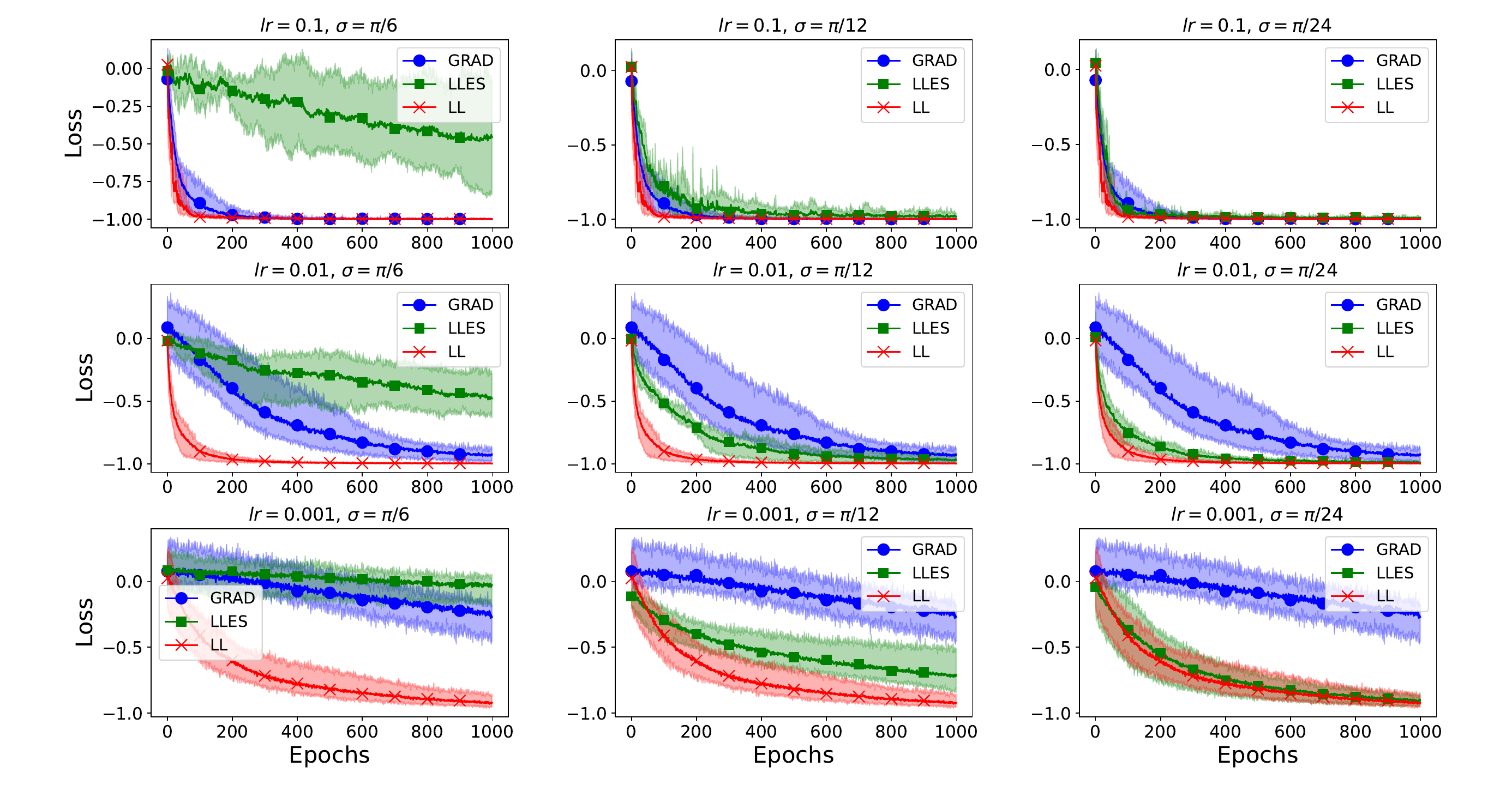}
    \caption{
    Cost function behavior during training using an 8-qubit quantum circuit with $L=8$, as illustrated in Fig. \ref{fig:model_circuit_gs}. Notably, for $lr=0.1$ and $\sigma=\pi/6,$ the LLES method failed to achieve the ground state. However, as $\sigma$ decreases, the method successfully reaches the ground state.}
    \label{fig:grafico_nq_8_nl_8_gs}
\end{figure*}

In Figure \ref{fig:grafico_nq_8_nl_8_gs}, we can observe the behavior of the cost function during training using the three optimization methods. In this case, we consider a quantum circuit with 8 qubits and $L=8$ layers. The results obtained for other configurations, i.e., quantum circuits with different numbers of qubits and depths $L$, are shown in Appendix \ref{apend:Additional_results}. To analyze how the initialization of the parameters \(\varphi\) influences the performance of the three methods, we ran the simulations 5 times for each configuration. This allows us to observe the average behavior of the cost function (represented by dark colors) and the maximum and minimum values (shaded area). Since we are using the PyTorch library to create our models, the parameters \(\varphi\) are automatically generated by PyTorch.

Additionally, as presented in Table \ref{tb:metodo_custo}, the computational cost of the LL and LLES methods scales linearly with $T$; in other words, the higher the value of $T$, the greater the computational cost. Considering that the primary goal of the LLES method is to minimize computational cost, we chose to use $T=2$, as this value proved sufficient to confirm the effectiveness of the LLES method while maintaining the lowest possible computational cost. However, we emphasize that, in theory, any value of $T$ could be employed.

From the results shown in this figure, we can make the following observations:

\begin{itemize} 
\item $lr=0.1$: The GRAD and LL methods were able to reach the ground state energy, and the initialization of the parameters had little influence on the results. On the other hand, the LLES method was unable to estimate the ground state energy when $\sigma = \pi/6$. However, as the value of $\sigma$ decreases, the LLES method converges to the ground state energy. Furthermore, the influence of the parameter initialization decreases as the value of $\sigma$ is reduced.

\item $lr=0.01$: In the GRAD method, it is observed that the cost function decreases as the training progresses, but now with a greater dependence on the initialization of the parameters \(\varphi\). In the LL method, the ground state energy is again reached, with no significant influence of the parameter initialization on the results. In the LLES method, we see that for $\sigma=\pi/6$, the cost function did not converge to the ground state energy. However, as $\sigma$ decreases, the method is able to optimize the parameters to achieve the desired result, and once again, the influence of the initialization decreases as $\sigma$ is reduced. 

\item  $lr=0.001$: Immediately, we observe that the GRAD method was unable to obtain the ground state energy. However, the LL method, although it also did not achieve the exact energy, obtained a value very close to it, indicating that treating the optimization rule as a learning problem indeed results in more efficient optimization of the parameters $\pmb{\theta}$. Finally, in the LLES method, we see a dependence on the choice of $\sigma$, where smaller values of $\sigma$ yield better results. For the smallest value of $\sigma$, the LLES method produces results similar to those obtained with the LL method.

\end{itemize}

In conclusion, the LLES method offers an efficient and versatile alternative for obtaining the ground state of quantum systems compared to traditional gradient-based methods. While it may not outperform the LL method under certain conditions, its computational efficiency and competitive performance make it an attractive choice, especially for complex systems where obtaining results with the LL method would be prohibitively expensive. 
Furthermore, as can be observed, the choice of $\sigma$ will influence the model's performance. An inappropriate choice, although resulting in lower computational cost, may prevent the model from converging to the minimum. In other words, the LLES method can balance computational efficiency and high-quality results, especially when hyperparameters are chosen appropriately.

\subsection{Quantum neural network classification}\label{subsec:qnn}

\begin{figure}
    \centering
    \includegraphics[scale=0.55]{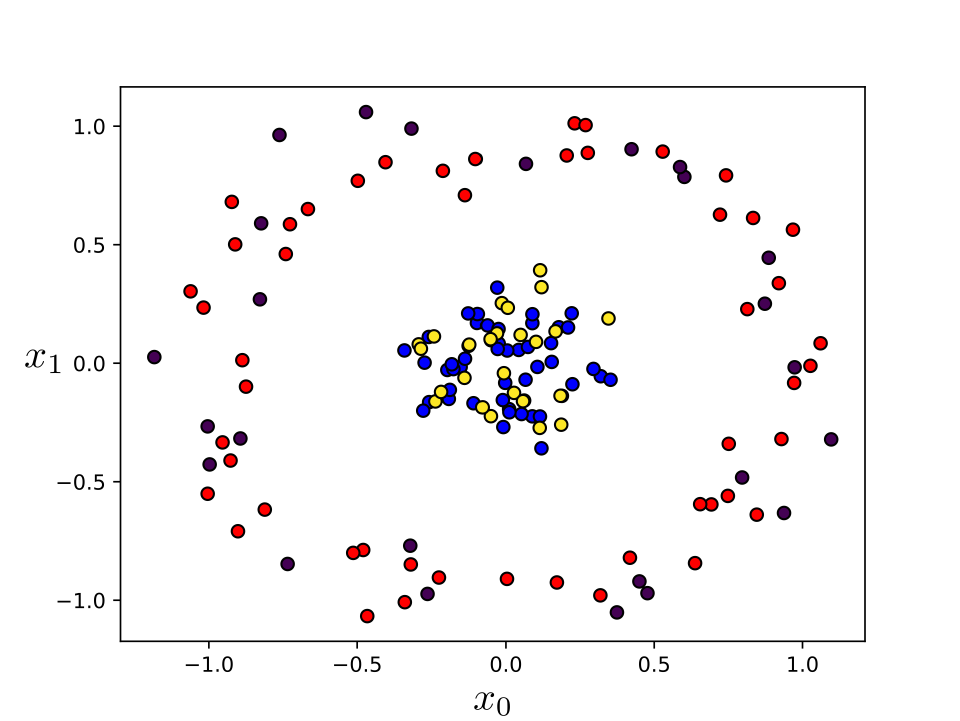}
    \caption{
    This figure shows the data used to train and test a quantum neural network. The data is divided into two sets, namely, the training data, points in red and blue, and the test data, points in purple and yellow. Each point is represented by two values $ \pmb{x}_{i} = (x_{0},x_{1})$ which are its coordinates on the Cartesian axis. For each point $i$ a label $y_{i} \in (0,1) $ is assigned.
    }
    \label{fig:data_classification}
\end{figure}

In this section, we will explore the application of the LLES method in training a quantum neural network for binary classification. We will utilize the dataset depicted in Fig. \ref{fig:data_classification}, which is split into two subsets: the training data, indicated by the red and blue circles, and the test data, represented by the purple and yellow circles. Each data point is associated with a binary label (0 or 1), and our objective is to train the model to accurately predict these labels.

\begin{figure}[t]
    \centering
    \includegraphics[scale=0.24]{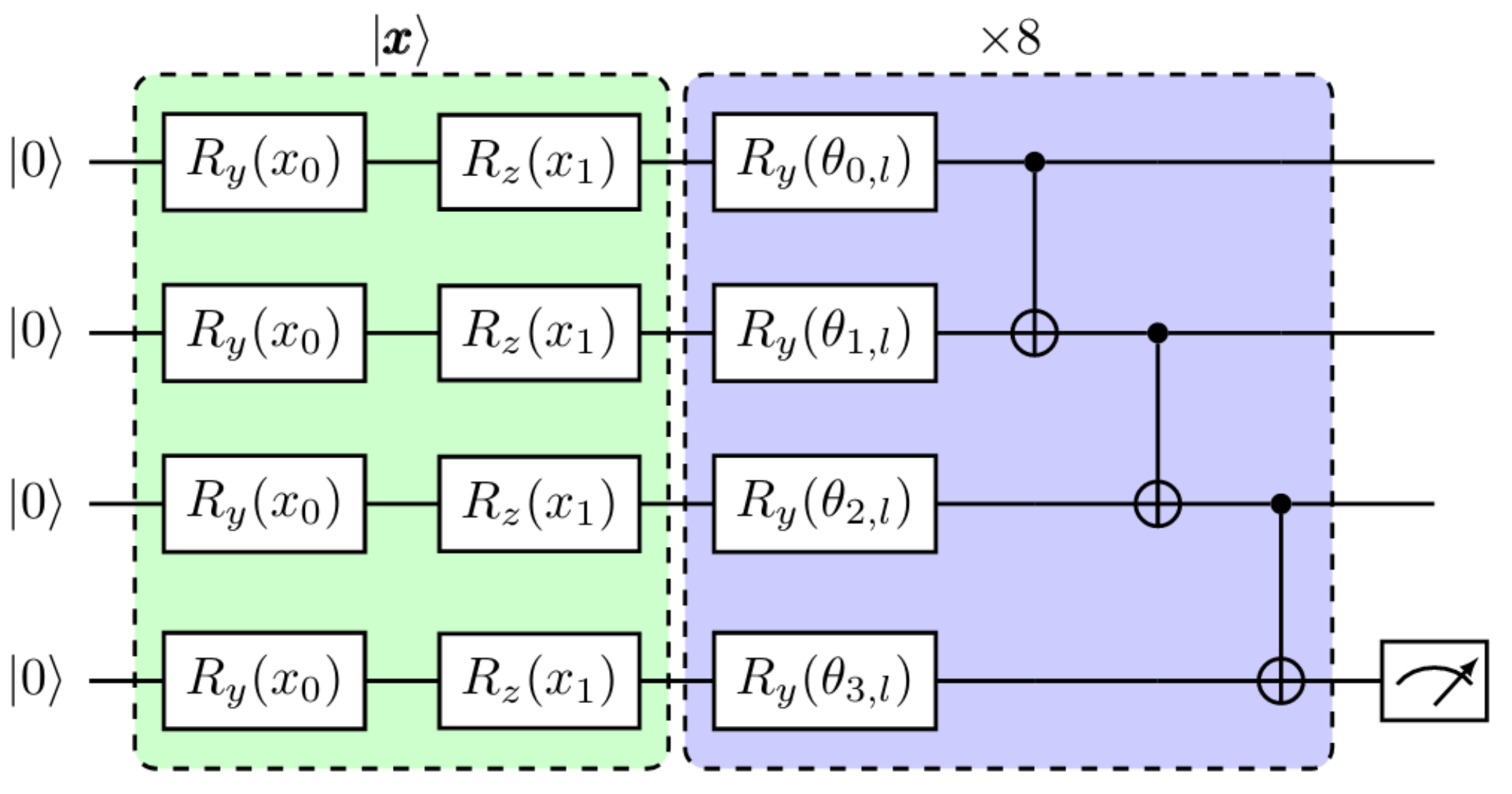}
    \caption{
    Illustration of the quantum circuit used for binary classification. Initially, data is encoded into a quantum state using a parameterization, highlighted in green in the illustration. Subsequently, this parameterized state undergoes transformations through a series of 8 layers, represented by $U_l$, highlighted in magenta in the illustration. The circuit concludes with a measurement of the last qubit to determine the probability of it being in the $|0\rangle$ state.}
    \label{fig:ilustracao_qnn}
\end{figure}

For this classification task, we will employ the quantum neural network depicted in Fig. \ref{fig:ilustracao_qnn}. The initial step involves encoding our data into a quantum state, which is achieved using the parameterization indicated in light green in Fig. \ref{fig:ilustracao_qnn}. Subsequently, we will utilize eight hidden layers, each denoted as $U_{l}$. These layers, as highlighted in light blue in Fig. \ref{fig:ilustracao_qnn}, are constructed by applying rotation gates $R_{y}$ to each qubit, followed by the application of CNOT gates between pairs of qubits. Finally, we will perform measurements on the last qubit, using the observable operator $O = |0 \rangle \langle 0|$.

We employ the Mean Squared Error (MSE) Loss function from the PyTorch library \cite{Paszke_pytorch} as our cost function $C_{j}$ in Eq. \eqref{eq:newLossFunction}. In scenarios where we use the LL and LLES methods, we utilize the mean of the MSE loss function as the cost function. Throughout the training process, we leverage the test data to assess how well the model can classify data that was not part of its training dataset. To evaluate this, we analyze the accuracy of the model.

\begin{figure*}
    \centering
    \includegraphics[scale=0.4]{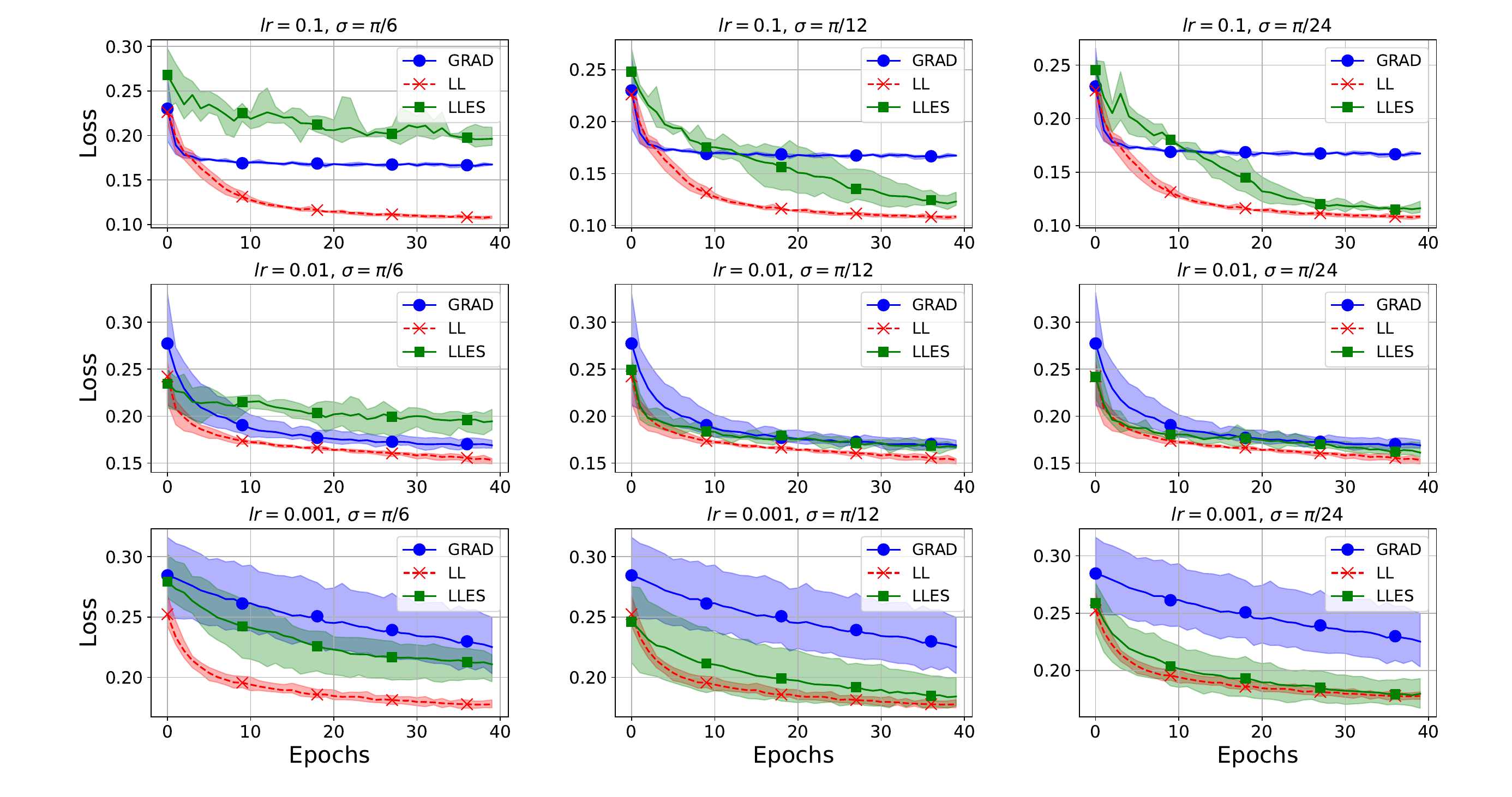}
    \caption{
    Training progress of the cost (loss) function for three different optimization methods using the average of the MSE loss function as the cost function. Notably, with a value of $\sigma = \pi/24$, the LLES method closely approximates the performance of the LL method, despite the LL method incurring higher computational costs.}
    \label{fig:grafico_loss_classificacao}
\end{figure*}

\begin{figure*}
    \centering
    \includegraphics[scale=0.4]{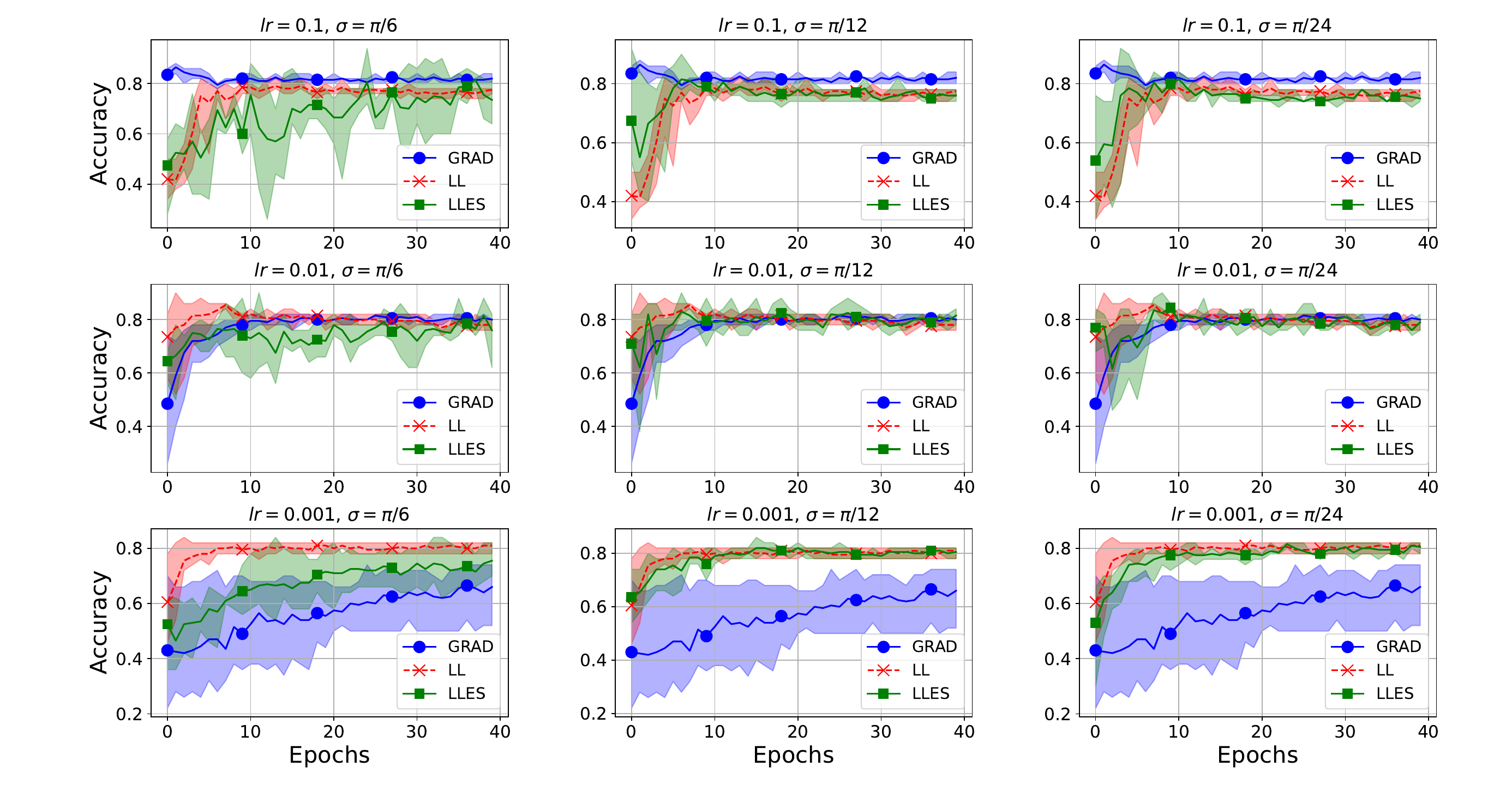}
    \caption{
    Accuracy evolution during training for all three optimization methods. Overall, the results obtained by the methods are quite similar. Notably, for a learning rate of 0.001, both the LL and LLES methods outperform the GRAD method in terms of accuracy.}
    \label{fig:grafico_acc_classificacao}
\end{figure*}

Similarly to the previous task, here we employ three learning rate values: $lr = 0.1$, $lr = 0.01$, and $lr = 0.001$. Additionally, we use three different values for $\sigma$: $\sigma = \pi/6$, $\sigma = \pi/12$, and $\sigma = \pi/24$. The organization of the graphs in Figures \ref{fig:grafico_loss_classificacao} and \ref{fig:grafico_acc_classificacao} is    analogous to that used for the previous task.

We commence by examining the behavior of the cost function during training, as illustrated in Figure \ref{fig:grafico_loss_classificacao}. In the case where we use $lr=0.1$, we observe that the cost function's behavior with the LLES method initially performed less favorably compared to the other two methods, particularly with $\sigma = \pi/6$. Nevertheless, as the value of $\sigma$ decreased, its behavior approached that of the LL method. When using $lr=0.01$, the three methods exhibited similar behavior, except for the case of $\sigma=\pi/6$, where the LLES method displayed relatively lower performance than the other two. However, once more, as the value of $\sigma$ decreased, the LLES method's behavior improved, approaching the performance of the LL method. Finally, for $lr=0.001$, initially, with $\sigma = \pi/6$, the cost function's behavior when using the LLES method fell between that of the GRAD and LL methods. However, its behavior approached the LL method as $\sigma$ decreased. Notably, in all cases, the LLES method appeared to be more influenced by parameter initialization compared to the LL method.


Fig. \ref{fig:grafico_acc_classificacao} shows the behavior of the model in classifying data that was not previously used to train it. This figure shows the accuracy of the model, that is, the mean accuracy. For this, we use that for a given input $x_{i}$ the label predicted by the model is given by
\begin{equation}
    y_{i} = \left\{ \begin{array}{rcl}
    1 & \mbox{if} &  \langle O \rangle  \geqslant  0.5,\\
    0 & \mbox{if} &  \langle O \rangle  < 0.5.\\
\end{array}\right.
\end{equation}

As seen in Figure \ref{fig:grafico_acc_classificacao}, for $lr=0.1$ the model trained using the GRAD method quickly converged to a maximum accuracy. However, this accuracy was suboptimal, as it did not reach the ideal value of 1. In contrast, the LL and LLES methods started with a lower accuracy and, during training, approached the accuracy achieved by the GRAD method. Notably, while the cost function values obtained using the LL and LLES methods were lower than those obtained with the GRAD method (as observed in Fig. \ref{fig:grafico_loss_classificacao}), the resulting accuracies were relatively similar.

With a learning rate of $lr=0.01$, all three methods showed similar behavior, as depicted in Figure \ref{fig:grafico_acc_classificacao}. However, for a learning rate of $lr=0.001$, it is evident that the model trained with the GRAD method exhibited inferior performance compared to the other two methods. Additionally, as the value of $\sigma$ decreased, the behavior of the model trained with the LLES method approached the behavior of the LL method.

Similar to the previous task, in general, the LL method achieved the best results for the binary classification problem. However, as seen earlier, when we consider the smaller values of $\sigma$, the results obtained with the LLES method closely resemble those of the LL method. Therefore, one can argue that the LLES method is superior to the LL method since it requires significantly fewer evaluations of the cost function while delivering comparable performance. Concerning the GRAD method, it is evident that the LLES method outperforms it for two primary reasons: it necessitates fewer cost function evaluations and provides equivalent or better results, especially when using a small learning rate of $0.001$.



\section{Conclusion}
\label{sec:Conclusion}

In this work, our primary objective was to introduce a novel optimization method, which we named `Learning to Learn with Evolutionary Strategy' (LLES). LLES represents the fusion of two well-established methods, Learning to Learn (LL) and Evolutionary Strategy (ES). Our motivation arises from the realization that relying solely on gradients for optimizing function parameters often neglects valuable information crucial to the optimization process. Notably, the gradient-based method tends to overlook crucial details such as entanglement levels in Variational Quantum Algorithms (VQAs). The LL method emerged as a powerful alternative, treating optimization as a task that can be learned itself. However, the computational burden of the LL method is substantial due to the large number of function evaluations required for parameter optimization.

In response to this challenge, we proposed employing the ES method for gradient estimation. We showed that ES can dramatically reduce the number of function evaluations compared to conventional gradient estimation techniques like the parameter shift rule.

To assess the effectiveness of the LLES method, we conducted experiments on two relevant quantum problems: ground state energy estimation for a given Hamiltonian and training a quantum neural network model. Our results indicated that LLES consistently outperformed or matched the performance of the LL and GRAD methods while requiring significantly fewer function evaluations. Moreover, we found that LLES's performance is profoundly influenced by the hyperparameter $\sigma$, where lower values tend to yield improved results.

While pinpointing an exact reason for the method's superior performance with smaller values of $\sigma$ poses a challenge, one possible explanation lies in our utilization of the evolutionary strategy to optimize LSTM layer parameters. This strategy provides a gradient estimate by generating a set of new parameters ($\pmb{\theta}_i$) using a normal distribution ($\mathcal{N}(\pmb{\theta},\sigma^2)$). By evaluating the cost function at these new parameters, we obtain an average behavior, effectively estimating the gradient based on the cost function behavior around $\pmb{\theta}$.

With a larger $\sigma$, the generated parameters ($\pmb{\theta}_i$) are further from $\pmb{\theta}$, impacting the accuracy of the estimate as we incorporate cost function behavior from distant points into the gradient estimation. Conversely, smaller $\sigma$ values result in parameters closer to $\pmb{\theta}$, enabling the gradient estimate to focus solely on the cost function's behavior around $\pmb{\theta}$, thus yielding a closer approximation to the true gradient value.
This explanation gains support from the observed similarity in the behavior of the LLES method with small $\sigma$ to that of the LL method, which explicitly employs the gradient.

Additionally, it is important to highlight a limitation related to the scalability of the proposed method. Currently, in the NISQ era, most advances in VQAs are achieved through numerical simulations performed on classical computers, which makes the analysis of larger systems particularly challenging. This technological limitation represents a significant obstacle to studying the scalability of VQAs, that is, evaluating how the method performs as the system size increases. In the case of this work, although the results obtained demonstrate that the proposed method is capable of producing satisfactory outcomes, its scalability remains an open question and requires further investigation.

In summary, this study introduces the LLES method as a promising addition to quantum optimization techniques. Its ability to achieve comparable or superior results while reducing computational overhead positions it as a valuable asset for quantum machine learning and optimization tasks. 

As a complement, in Appendices \ref{appendixA} and \ref{appendixB}, we investigate the applicability of the LLES method to more complex quantum problems. Specifically, we analyze the performance of this method in a noisy circuit, comparing it with the GRAD and LL methods, as well as its applicability to a quantum machine learning problem, focusing on a multiclass classification task. The results suggest that, as quantum computing technology advances, innovative optimization methods like LLES have the potential to play a fundamental role in realizing the full potential of quantum computing for practical applications.

\vspace{0.3cm}

\begin{acknowledgments}
This work was supported by the Coordination for the Improvement of Higher Education Personnel (CAPES) under Grant No. 88887.829212/2023-00, the National Council for Scientific and Technological Development (CNPq) under Grants No. 309862/2021-3, No. 409673/2022-6, and No. 421792/2022-1, and the National Institute for the Science and Technology of Quantum Information (INCT-IQ) under Grant No. 465469/2014-0.
\end{acknowledgments}

\vspace{0.1cm}

\textbf{Data availability.}
The numerical data generated in this work and code is avalilable at \burl{https://github.com/lucasfriedrich97/LLES}.

\vspace{0.1cm}

\textbf{Contributions} The project was conceived by L.F., who also carried out the simulations. J.M. supervised the research. L.F. wrote the first version of the article, which was revised by J.M..

\vspace{0.1cm}
\textbf{Competing interests.}
The authors declare no competing interests.


\appendix

\section{Additional results}
\label{apend:Additional_results}

In this appendix, we present additional results related to subsection \ref{subsec:Ground_state}. Specifically, we analyze the behavior of the cost function using the three optimization methods considered in this work for different configurations, i.e., varying numbers of qubits and parameterization depth \(L\). The quantum circuit used to obtain these results is similar to the one presented in Figure \ref{fig:model_circuit_gs}.

\begin{figure*}[]
    \centering
    \includegraphics[scale=0.39]{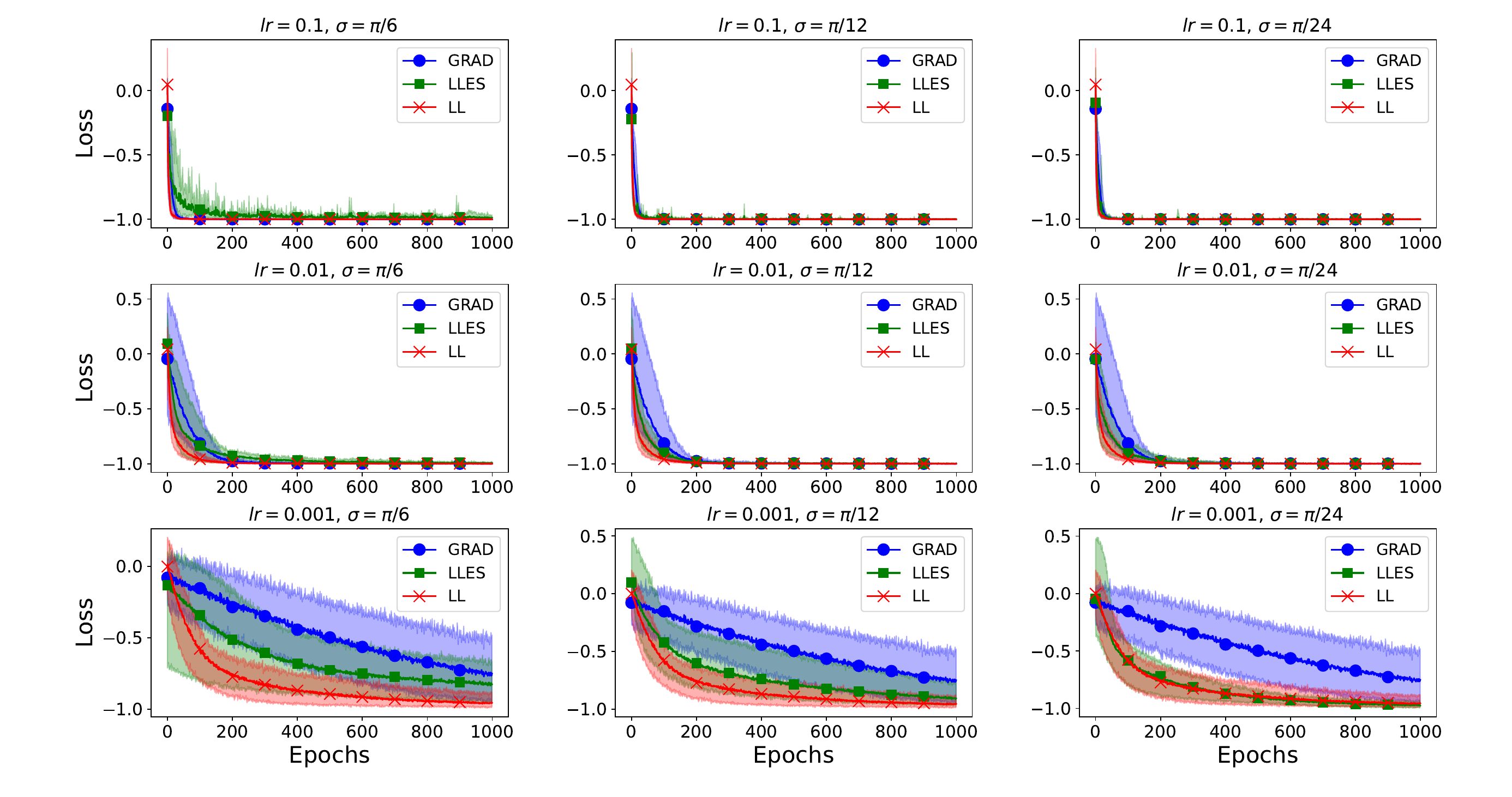}
    \caption{
    Performance comparison of three optimization methods in terms of the cost (loss) function during training. The three methods include the GRAD method, which relies on derivatives obtained using Eq. \eqref{eq:derivada_C}; the Learning to Learn (LL) method, also utilizing gradients from Eq. \eqref{eq:derivada_C}; and the Learning to Learn with an Evolutionary Strategy (LLES) method, which estimates gradients using Eq. \eqref{eq:es_3}. The results are obtained using a quantum circuit with 4 qubits and $L=4$, as illustrated in Fig. \ref{fig:model_circuit_gs}. The influence of different learning rates ($lr$) and LSTM parameter values ($\sigma$) on model performance is analyzed. The three rows correspond to $lr=0.1$, $lr=0.01$, and $lr=0.001$, respectively, while the three columns correspond to $\sigma = \frac{\pi}{6}$, $\sigma = \frac{\pi}{12}$, and $\sigma = \frac{\pi}{24}$, respectively. Each configuration is executed five times to assess the impact of parameter initialization, and the results are depicted with dark lines representing the average values, with dashed regions indicating the maximum and minimum value variations.}
    \label{fig:groundState_fig_1}
\end{figure*}

\begin{figure*}[]
    \centering
    \includegraphics[scale=0.39]{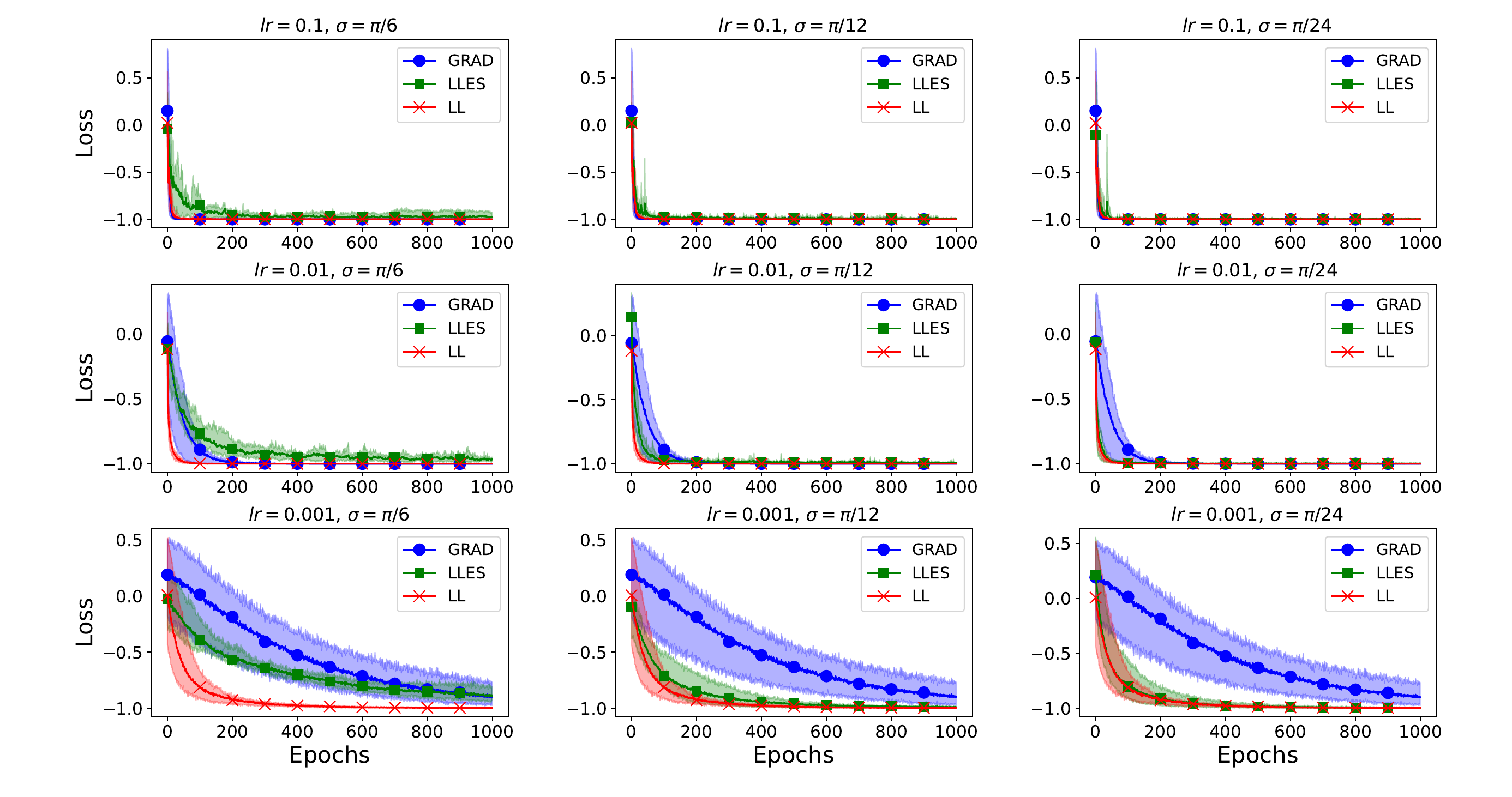}
    \caption{
    Similarly to Fig. \ref{fig:groundState_fig_1}, these graphs depict the cost (loss) function behavior during training, but now the quantum circuit comprises 4 qubits with $L=8$. The chosen values of $lr$ and $\sigma$ are the same as those used to obtain the results shown in Fig. \ref{fig:groundState_fig_1}. As before, the dark lines represent the average behavior of the three methods across five simulations, while the lighter-shaded areas indicate the range between the maximum and minimum values among the five simulations.}
    \label{fig:groundState_fig_2}
\end{figure*}

\begin{figure*}[]
    \centering
    \includegraphics[scale=0.39]{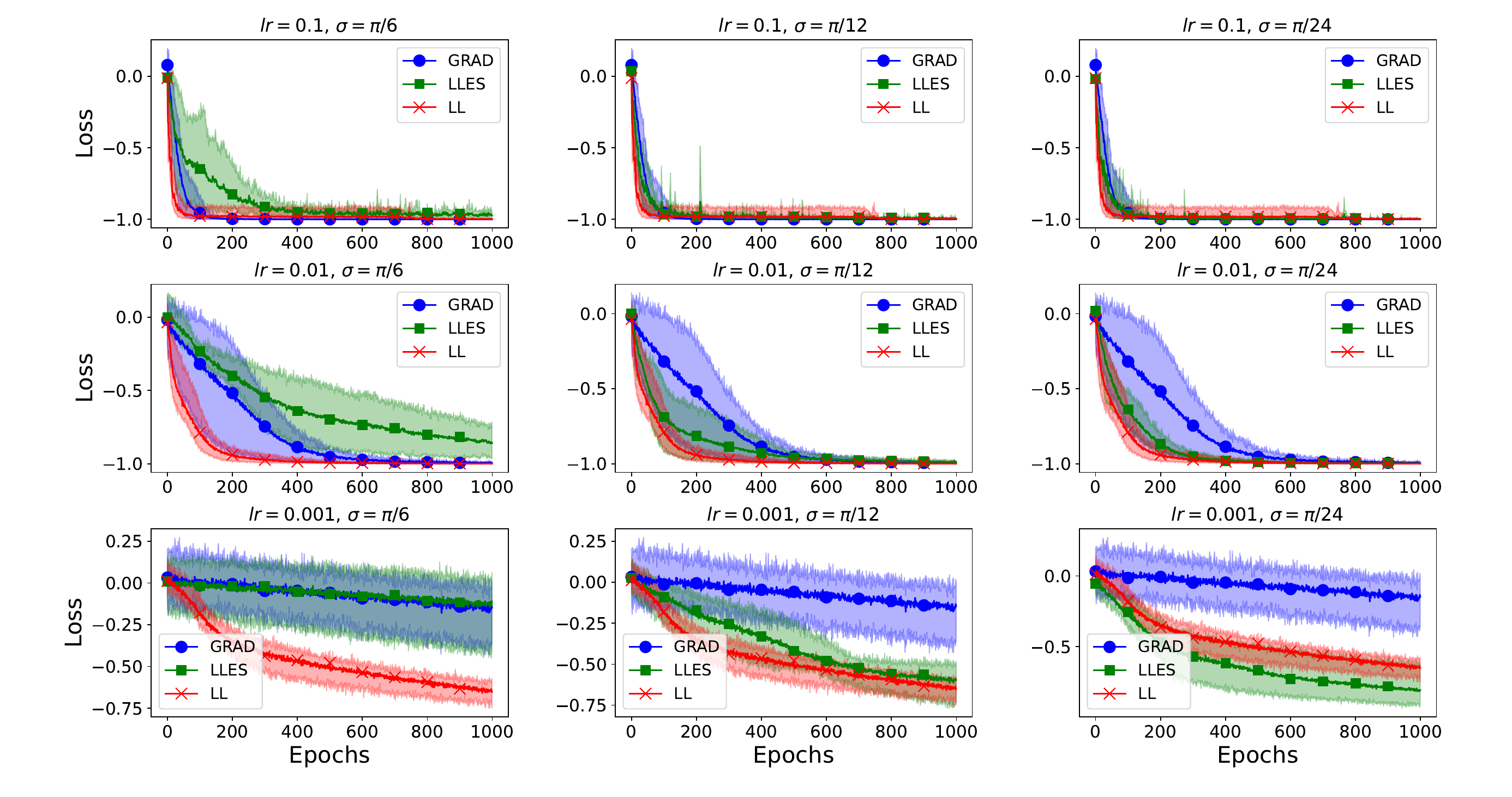}
    \caption{
    Cost function behavior during training using an 8-qubit quantum circuit with $L=4$, as depicted in Fig. \ref{fig:model_circuit_gs}. Notably, when $lr=0.001$, none of the methods succeeded in reaching the ground state, for the number of epochs used. Additionally, for $\sigma=\pi/24$, the cost function behavior using the LLES method outperformed the other two methods.}
    \label{fig:groundState_fig_3}
\end{figure*}

The figures \ref{fig:groundState_fig_1}, \ref{fig:groundState_fig_2}, and \ref{fig:groundState_fig_3} illustrate the results obtained using quantum circuits with 4 qubits and \(L=4\); 4 qubits and \(L=8\); and 8 qubits and \(L=4\), respectively. As in the case discussed in the main text, to analyze the influence of parameter initialization \(\varphi\) on the three methods, simulations were repeated five times for each configuration. The dark lines, highlighted by markers, represent the average behavior of the cost function, while the shaded regions indicate the observed maximum and minimum behaviors. Additionally, for these results, we considered \(T=2\).

The obtained results corroborate the conclusions presented in the main text. In particular, we observe that, by structuring the optimization rule as a learning process, the model becomes more efficient in optimizing the parameters \(\pmb{\theta}\). Furthermore, it is evident that the LLES method is capable of achieving results analogous to those obtained with the LL method, which, as demonstrated in Table \ref{tb:metodo_custo}, is more computationally expensive. However, these analogous results depend on the appropriate choice of hyperparameters, particularly the hyperparameter \(\sigma\).

\section{Application to ground state energy estimation}
\label{appendixA}

\begin{figure}[]
    \centering    \includegraphics[scale=0.3]{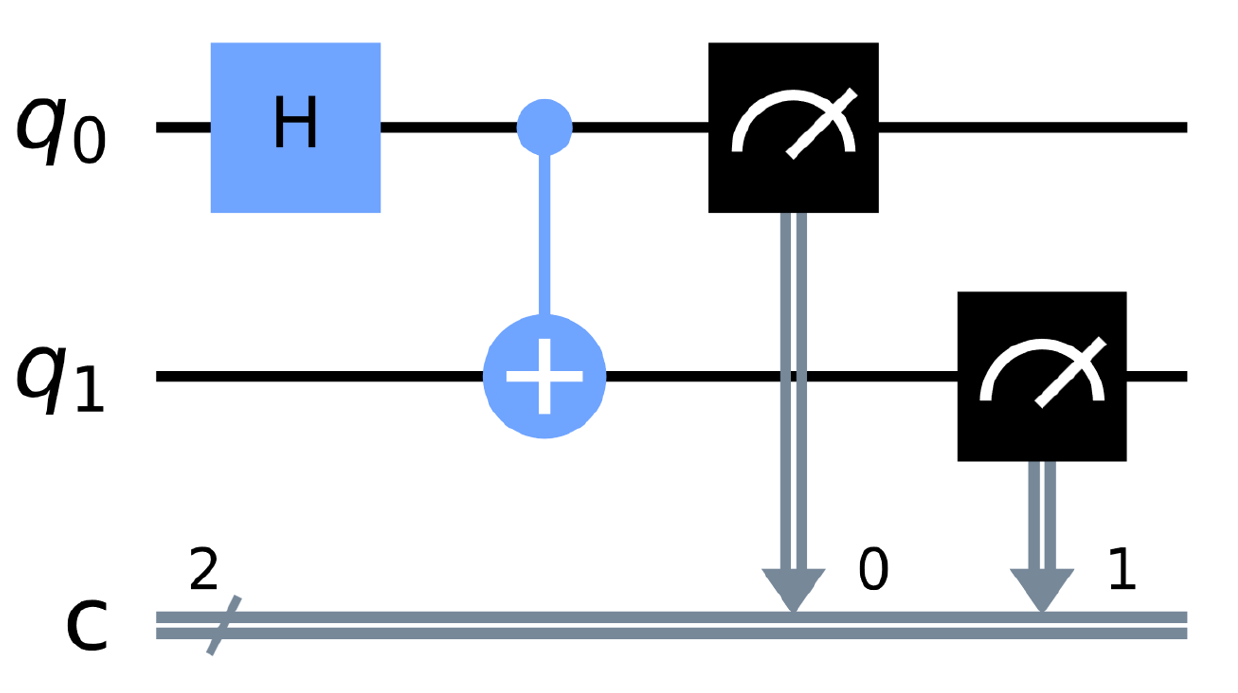} \caption{Quantum circuit used to prepare the Bell state $|\Phi^{+} \rangle$.}
    \label{fig:circuitRef}
\end{figure}

In this appendix, we investigate the impact of noise on our new method, aiming to assess its behavior on a real quantum computer. To conduct this analysis, we will undertake simulations akin to those employed to derive the results presented in Fig. \ref{fig:grafico_nq_8_nl_8_gs}.  That is, we consider a ground state estimation problem, where the Hamiltonian is given by
\begin{equation}
     H = \bigotimes_{j=1}^{n} \sigma_{Z}^{j},
\end{equation}
and the parametrization is given by Fig. \ref{fig:model_circuit_gs}. However, we will now introduce the correlated amplitude damping error, a form of noise, into the simulations. The degree of influence exerted by this noise will be characterized by a parameter $\lambda \in (0,1)$, where higher values signify a more pronounced effect. We emphasize that the $\lambda$ used in this appendix to describe the noise intensity is different from the $\lambda$ used in the main text where it is associated with the ES method. To illustrate the impact of this noise, consider the quantum circuit depicted in Figure \ref{fig:circuitRef}, utilized to prepare the following Bell state:
\begin{equation}
    |\Phi^{+} \rangle = \frac{1}{\sqrt{2}}( |00\rangle + |11\rangle ).
\end{equation}

Hence, upon constructing and executing this circuit, we anticipate the following probabilities: $P(|00\rangle,|\Phi^+\rangle) = 0.5$ and $P(|11\rangle,|\Phi^+\rangle) = 0.5$. This indicates that in half of the circuit executions and subsequent measurements, we observe the state $|00\rangle$, while in the remaining half, we observe the state $|11\rangle$.

\begin{figure}[]
    \centering    \includegraphics[scale=0.5]{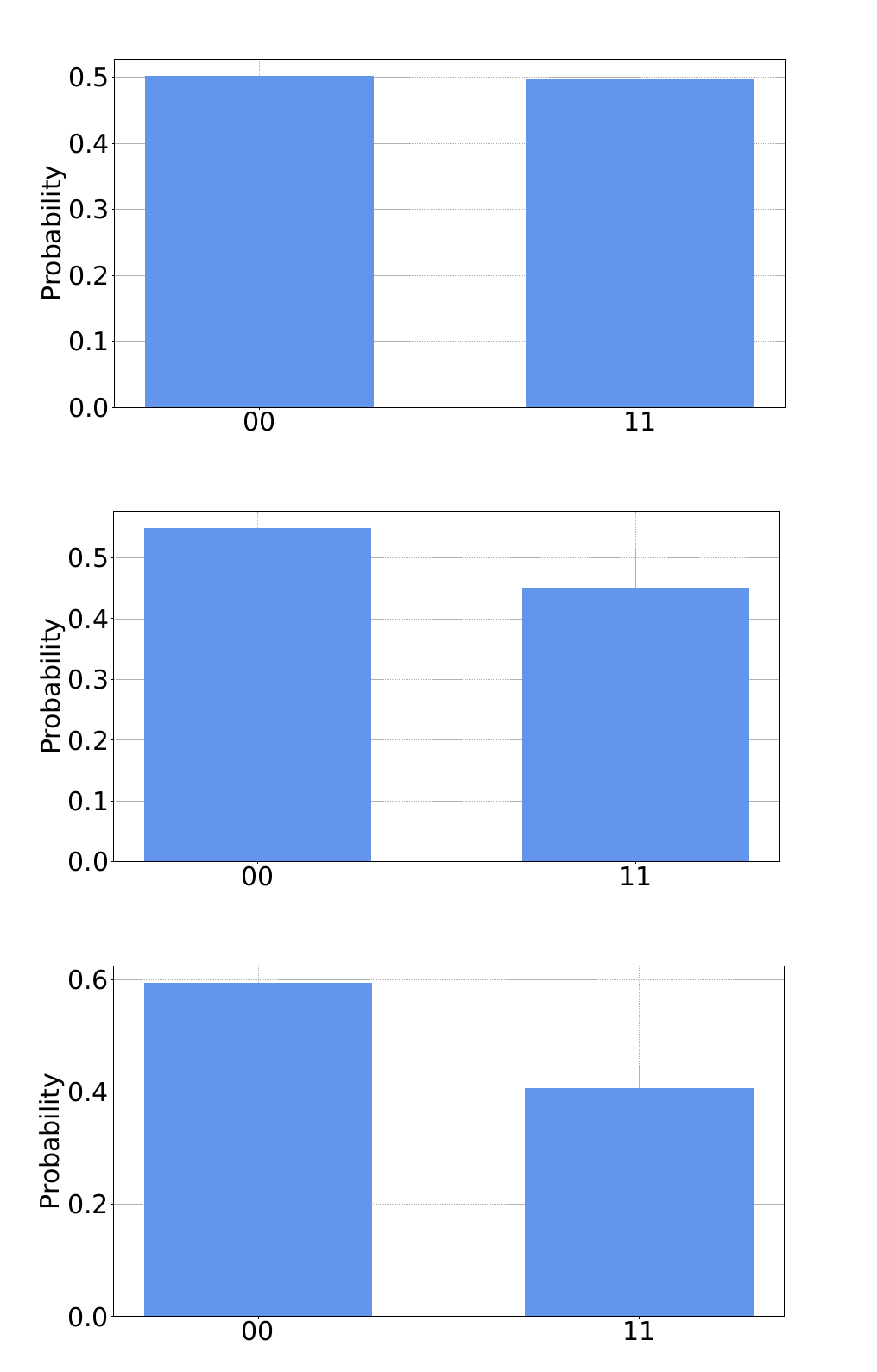}   \caption{In this figure, we show the impact of varying the noise parameter $\lambda$, which signifies the extent of influence of the amplitude damping error noise, on the probabilities associated with the Bell state $|\Phi^{+} \rangle$.
We begin by examining the probabilities when $\lambda=0$ (top graph), followed by an assessment of how these probabilities change when $\lambda=0.1$ (middle graph). Finally, we analyze the behavior of the probabilities associated with the state $|\Phi^{+} \rangle$ when $\lambda=0.2$ (bottom graph).}
    \label{fig:noiseRef}
\end{figure}

In Fig. \ref{fig:noiseRef}, we observe the influence of varying the noise parameter $\lambda$ on the results. Initially, in the top part of the figure, with $\lambda=0$, the probabilities of obtaining states $|00\rangle$ and $|11\rangle$ are close to $0.5$, as anticipated. However, due to statistical errors inherent in the model, these probabilities deviate slightly from the ideal value of $0.5$. As we increase $\lambda$ to $0.1$, as shown in the middle part of Fig. \ref{fig:noiseRef}, we witness a rise in the probability of obtaining state $|00\rangle$, accompanied by a decrease in the probability of obtaining state $|11\rangle$. Finally, for $\lambda=0.2$, depicted in the bottom part of Fig. \ref{fig:noiseRef}, we observe that the probability associated with state $|00\rangle$ approaches $0.6$, while the probability associated with state $|11\rangle$ approaches $0.4$.

The outcomes depicted in Fig. \ref{fig:noiseRef} suggest that the amplitude damping error, acting as noise, simulates the loss of energy from the system to the environment. This phenomenon occurs because as the system loses energy to the environment, the likelihood of observing the state $|00\rangle$, associated with lower energy, increases.

In our subsequent analysis, we will explore the behavior of the new method in the presence of noise. As illustrated in the results presented in the main text, the selection of the hyperparameter $\sigma$ holds significant importance for the effectiveness of the proposed method. Notably, for the cases examined, the optimal value was determined to be $\sigma=\pi/24$. Therefore, in the forthcoming results, we will exclusively utilize this value.

This choice is substantiated by our objective in hyperparameter selection, which is to achieve optimal results. Thus, given our prior identification that the best outcomes are attained with small values of $\sigma$, our analysis will concentrate solely on evaluating the impact of noise on these pertinent cases.

\begin{figure}[h]
    \centering
    \includegraphics[scale=0.55]{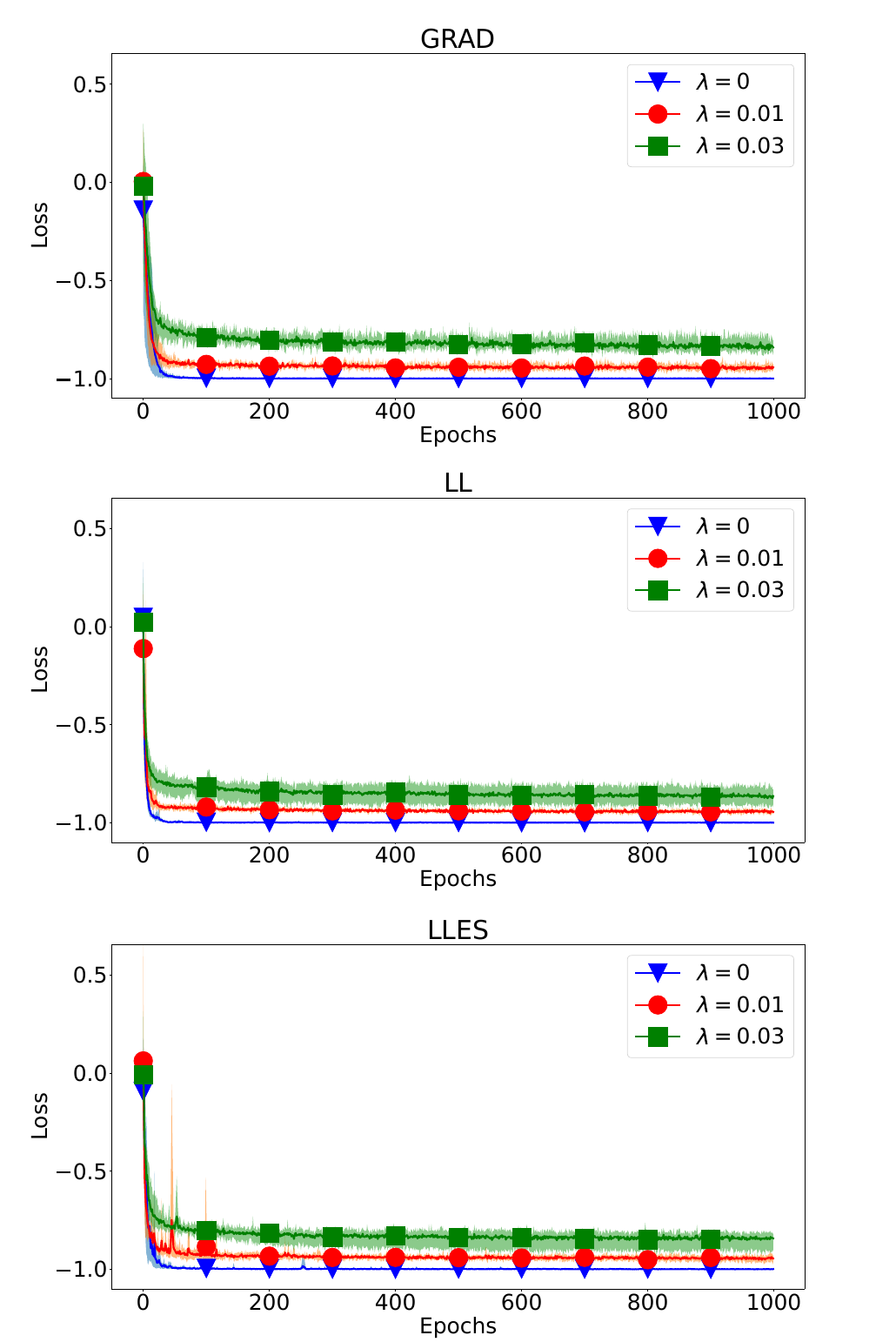}
    \caption{
    The loss function is plotted against the number of epochs for three distinct values of the noise parameter $\lambda$, while maintaining a constant learning rate of $lr=0.1$. The top figure corresponds to the GRAD method, the middle figure depicts the LL method, and the bottom figure illustrates the newly proposed LLES method.}
    \label{fig:noise_0.1}
\end{figure}

\begin{figure}[h]
    \centering
    \includegraphics[scale=0.55]{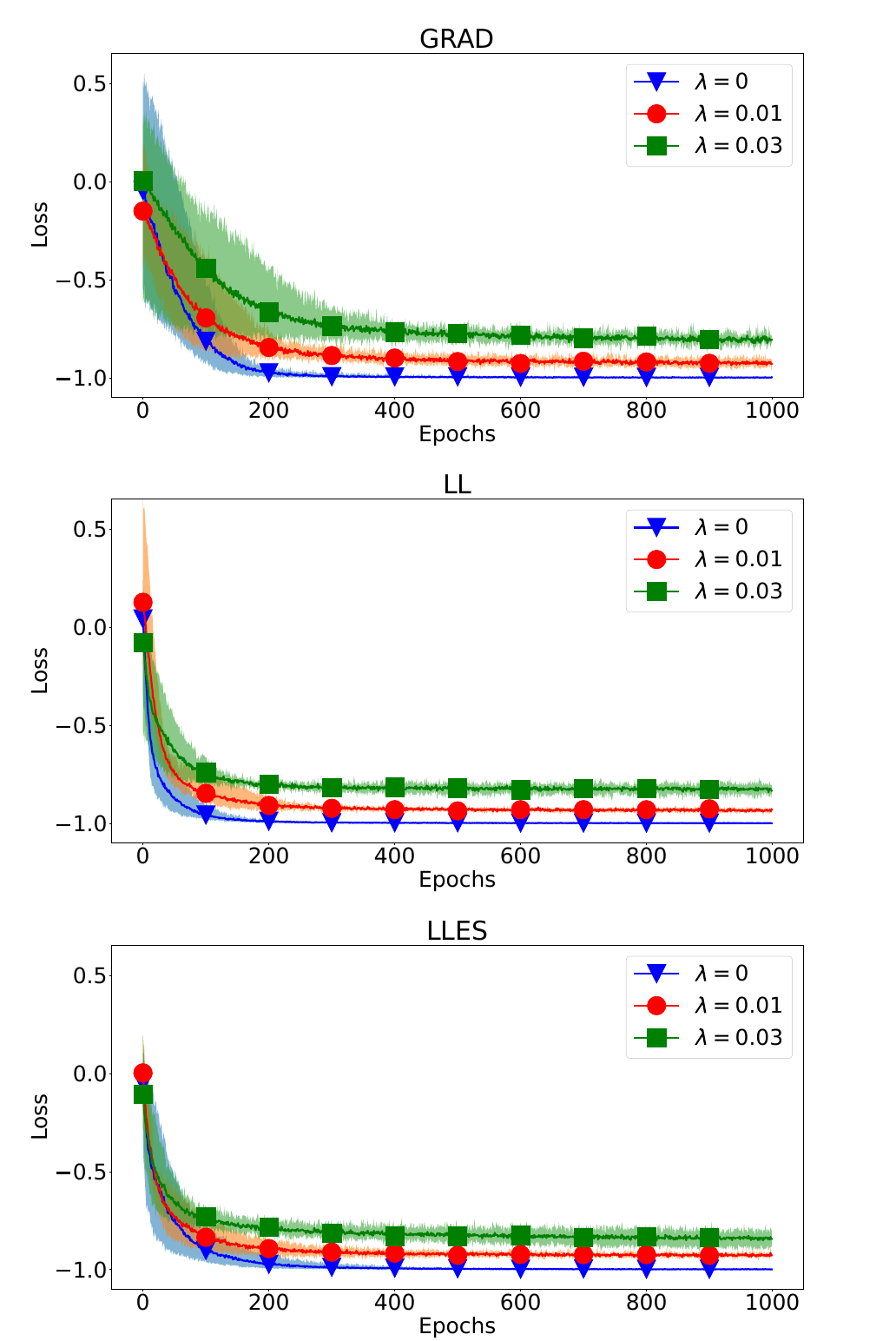}
    \caption{
    The loss function is plotted against the number of epochs for three distinct values of the noise parameter $\lambda$, while maintaining a constant learning rate of $lr=0.01$. The top figure corresponds to the GRAD method, the middle figure depicts the LL method, and the bottom figure illustrates the newly proposed LLES method.}
    \label{fig:noise_0.01}
\end{figure}

\begin{figure}[h]
    \centering
    \includegraphics[scale=0.55]{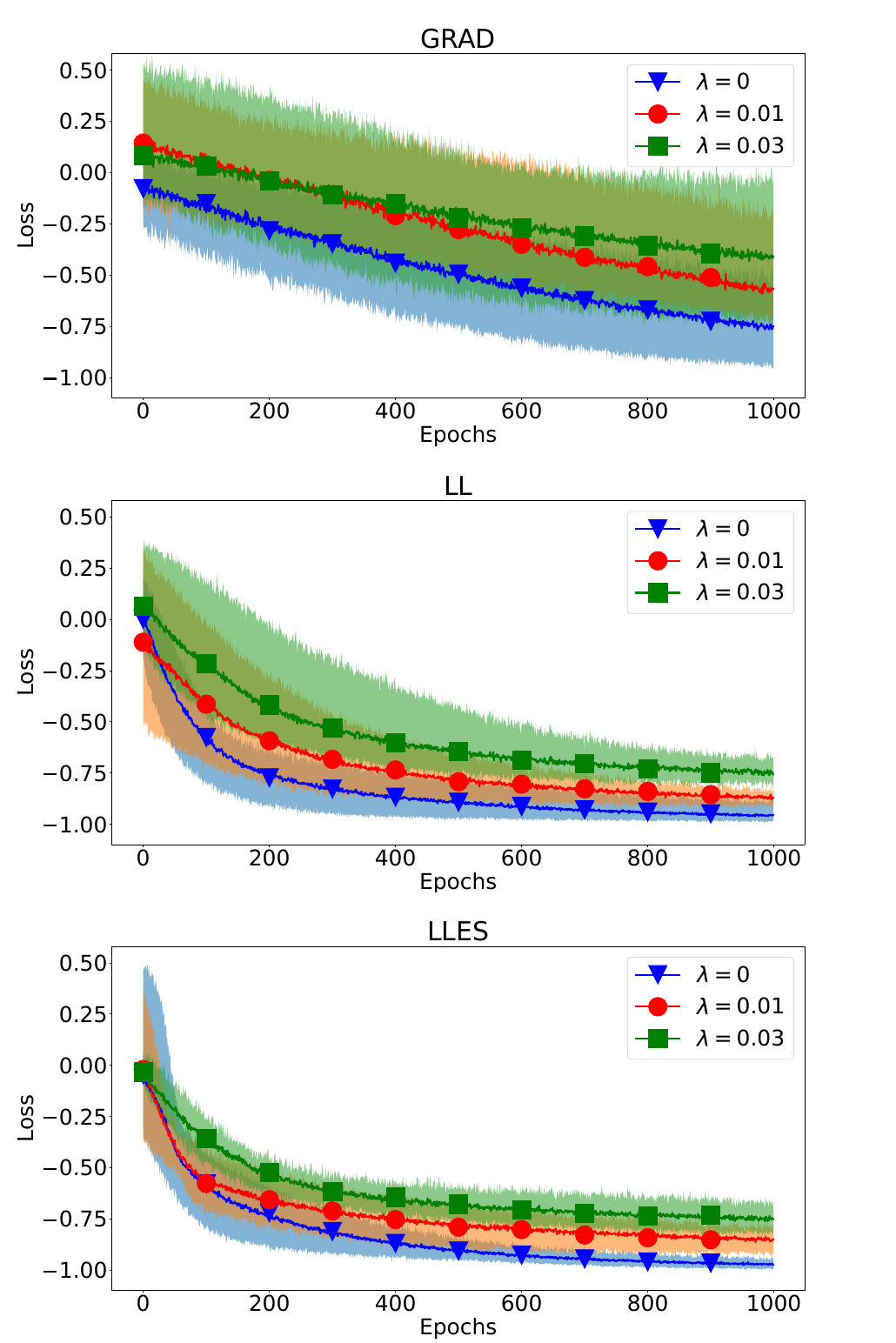}
    \caption{
    The loss function is plotted against the number of epochs for three distinct values of the noise parameter $\lambda$, while maintaining a constant learning rate of $lr=0.001$. The top figure corresponds to the GRAD method, the middle figure depicts the LL method, and the bottom figure illustrates the newly proposed LLES method.}
    \label{fig:noise_0.001}
\end{figure}

From the findings depicted in Figs. \ref{fig:noise_0.1}, \ref{fig:noise_0.01}, and \ref{fig:noise_0.001}, it's evident that all three methods exhibit similar trends. As the parameter $\lambda$ increases, each method demonstrates the ability to optimize and converge towards a minimum value for $\langle H \rangle$. However, for $\lambda \neq 0$, none of the methods are able to attain the anticipated minimum value of $\langle H \rangle = -1$.

This observation aligns with the presence of noise in the system, causing energy loss to the environment. Consequently, it's reasonable to conclude that exact attainment of the $\langle H \rangle$ value is not possible due to this inherent noise-induced energy loss.

Hence, based on the obtained results, there's an indication that this new method, when implemented on a real quantum computer, could estimate a minimum value for $\langle H \rangle$ with reduced computational overhead compared to alternative methods. However, it's crucial to acknowledge that in the context of utilizing a real quantum computer, the minimum value derived by the LLES method, as well as by other methods, might deviate from the actual global minimum value.

This discrepancy arises because, as demonstrated in Fig. \ref{fig:noiseRef}, noise significantly influences the probabilities we obtain, consequently impacting the calculated value of $\langle H \rangle$. Therefore, while the method may offer computational advantages, its output on a real quantum computer remains subject to the inherent limitations imposed by noise.

\clearpage

\section{Application to quantum neural network training}
\label{appendixB}

In this section, we present the application of this new method to solve a multiclass classification problem. To do this, we use the MNIST dataset, which consists of a collection of handwritten digits widely used by the machine learning community. This dataset was used to create the dataset used in our experiments.

For the experiments, we used a dataset derived from MNIST, restricting the data to the digits zero, one and two. This allowed us to transform the original multiclass classification problem with 10 classes into a problem with just 3 classes. Furthermore, to avoid class imbalance issues, i.e., situations where the number of training examples for each class is unequal, we ensured that each class had the same amount of data. Thus, we built a training dataset containing 3000 examples, with 1000 examples for each of the three classes. Likewise, we created a test dataset with 300 examples, distributed equally among the three classes, with 100 examples per class.

The parameterization model used in these experiments is similar to the model shown in Fig. \ref{fig:model_circuit_gs}, represented in dark blue. However, to encode our data into a quantum state, we use the encoding scheme known as amplitude encoder. Furthermore, our model was created with 10 qubits, and the parameterization depth was set to $L=15$. For the measurements, we chose to individually measure the last three qubits of the model using the $H_{i} = |0\rangle \langle0|$ observable. Here the subindex $i$ indicates which qubit the observable will be applied to, for example, if $i=1$ then it will be applied to qubit $7$, if $i=2$ it will be applied to qubit $8$. For the cost function, we use the standard model is defined in Eq. \eqref{eq:costFunctionAppendixB}. This cost function,
\begin{equation}
    C(x,y) = \frac{1}{N}\sum_{l=1}^{N}\frac{1}{3}\sum_{p=1}^{3}( Tr[H_{p}U \rho(x_{l})U^{\dagger}] - y_{l,p})^2,\label{eq:costFunctionAppendixB}
\end{equation}
is known as the mean square error. In general, this cost function is used in regression problems, but we choose this cost function here duo to its simplicity and interpretation regarding accuracy.

In the case of the new method, we define the cost function as
\begin{equation}
    L = \frac{1}{T}\sum_{q=1}^{T}w_{q}C(x,y)_{q}, 
\end{equation}
with $C(x,y)_{q}$ being the cost function in Eq. \eqref{eq:costFunctionAppendixB}, in relation to the $q$-th layer, and $w_{q}$ is a parameter which defines the influence of the $q$-th $C(x,y)_{q}$ in this sum. In this case, we choose $w$ as variable.  In orther words, $w_{q} \neq w_{q+1} $. As we consider $T=2$, we use $w = [0.09090909090909091, 0.9090909090909091]$.
To analyze the influence of parameter initialization, we repeated the experiments 5 times for each configuration. Below we present the results obtained.

\begin{figure}[h]
    \centering
    \includegraphics[scale=0.5]{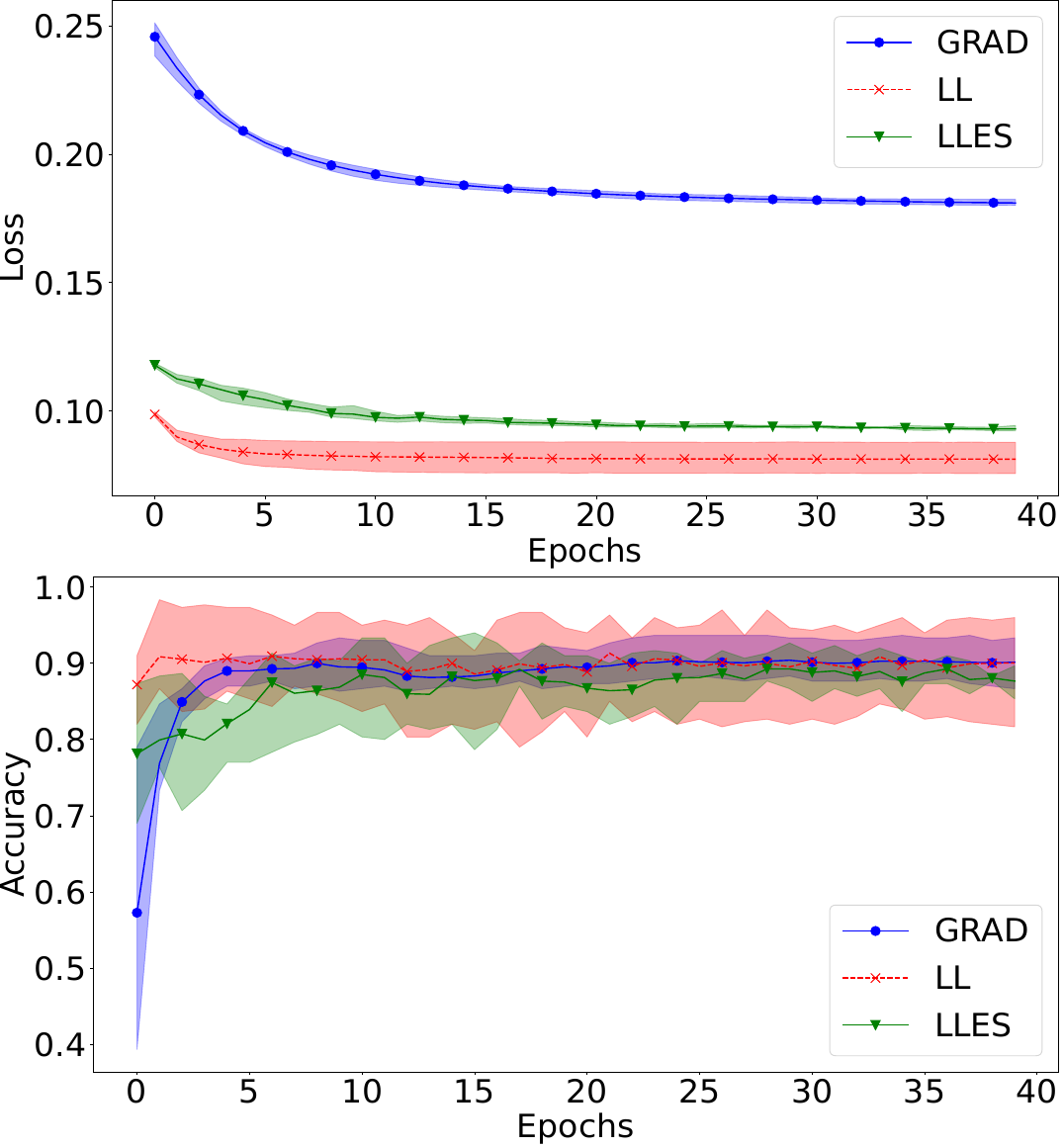}
    \caption{Cost function (upper graph) and accuracy (lower graph) as a function of the number of training epochs, applying the GRAD, LL, and LLES mehtods. In this case, we chose to use $\sigma=\pi/24$ in the LLES method.}
    \label{fig:MNISTresult}
\end{figure}

Based on the results shown in Fig. \ref{fig:MNISTresult}, we can make the following observations:
\begin{enumerate}
    \item GRAD Method:
     \begin{itemize}
         \item The behavior of the cost function starts around 0.25 and tends to 0.2 as training progresses. The initialization of the parameters did not significantly influence the behavior of the cost function, indicating that the GRAD method is robust with respect to parameter initialization. In other words, regardless of the parameter initialization, we obtain the same result using the GRAD method.
         \item However, the accuracy using the GRAD method shows a dependency on parameter initialization. This means that if our goal is to achieve good accuracy, it will depend on the initialization of the parameters when using the GRAD method.
     \end{itemize}
     \item LL Method:
     \begin{itemize}
         \item The cost function is significantly lower than the value obtained by the GRAD method. This indicates that if our main objective is to minimize the cost function, this method is more effective than the GRAD method.
         \item The initialization of the parameters has a greater influence on this method, suggesting that different results can be obtained depending on the initial parameters.
         \item The accuracy obtained using the LL method also depends on the initialization of the parameters and shows oscillations around 0.9.
     \end{itemize}

     \item LLES Method:
     \begin{itemize}
         \item The cost function starts at a value lower than the GRAD method but higher than the LL method. However, unlike the LL method, the initialization of the parameters did not influence the results in this case, indicating that the LLES method is more robust with respect to parameter initialization.
         \item Regarding accuracy, we observe a dependency on parameter initialization, similar to the other methods. Additionally, accuracy tends to approach 0.9 as training progresses, as with the other methods.
     \end{itemize}
\end{enumerate}

Based on the described results, we can argue that the LLES method is superior for the following reasons:
\begin{itemize}
    \item \textbf{Cost Function}: The LLES method achieves a lower cost function than the GRAD method, making it a more appropriate choice if the goal is to minimize the cost function. Although the cost function obtained with the LLES method is higher than that obtained with the LL method, it demonstrates greater robustness with respect to parameter initialization. Unlike the LL method, parameter initialization does not significantly influence LLES results, reducing the need for fine-tuning initial parameters and ensuring consistent results.
    \item \textbf{Accuracy}: All methods, unlike what is observed with the cost function, are influenced by parameter initialization and tend to achieve very similar results in terms of accuracy. However, since the computational cost associated with the LLES method is lower than that of the other methods, it is more recommended, despite its slightly lower accuracy.
    \item  \textbf{Computational Efficiency}: In general, the results obtained by the LLES method are close to those obtained by the LL method and better than those of the GRAD method when analyzing the cost function. In terms of accuracy, LLES results are comparable to those of the GRAD method, though slightly inferior to the LL method. However, the computational cost of the LLES method is significantly lower than that of the other methods. Therefore, considering the computational cost, the LLES method is preferable. It offers slightly lower accuracy but with a low computational cost, which is more advantageous than achieving slightly higher accuracy with a high computational cost.
\end{itemize}

Thus, considering the combination of robustness, efficiency, and computational cost, the LLES method stands out as the most recommended choice among the analyzed methods.


\end{document}